\documentclass[journal]{vgtc}                          

\graphicspath{{figures/}{pictures/}{images/}{./}} 

\usepackage{times}                     

\usepackage{tabu}                      
\usepackage{booktabs}                  
\usepackage{lipsum}                    
\usepackage{mwe}                       

\usepackage{mathptmx}                  

\usepackage{enumitem}
\usepackage{booktabs}
\usepackage{multirow}

\usepackage{colortbl}

\newcommand{\Note}[1] 
{\textcolor{black}{#1}}
\newcommand{\HDY}[1] 
{\textcolor{black}{#1}}
\newcommand{\ic}[1] 
{\textcolor{black}{#1}}
\newcommand{\KHC}[1]
{\textcolor{black}{#1}}

\newcommand{\nonsig}[1] 
{\textcolor{gray!70}{#1}}%

\newcolumntype{G}{>{\columncolor{gray!15}[\tabcolsep][\tabcolsep]\centering\arraybackslash}c}

\newcommand{\ciValues}[1] 
{\small{#1}\normalsize{}}
\usepackage{caption}
\usepackage{makecell}
\usepackage[normalem]{ulem}
\usepackage{booktabs}

\onlineid{0}

\vgtccategory{Research}

\title{Exploring the Effects of Olfactory Cues and Ventilation on Teleportation-based Navigation in VR}

\author{%
  \authororcid{Dongyun Han}{0000-0002-9517-5326},
  Heecheol Kim, 
  \authororcid{Siyeon Bak}{0009-0008-8845-2269},   \authororcid{Chang-Guen Song}{0000-0003-0204-6160}, \authororcid{Sun-Jeong Kim}{0000-0002-8663-4578},    and 
  \authororcid{Isaac Cho}{0000-0003-1582-8428}
}

\authorfooter{
  \item
  	Dongyun Han is with Clemson University.
  	E-mail: dongyuh@clemson.edu
   \item
Heecheol Kim, Siyeon Bak, Chang-Guen Song and Sun-Jeong Kim are with Hallym University
E-mail: khc3967@icloud.com, 123siyeon84@gmail.com, cgsong@hallym.ac.kr, sunkim@hallym.ac.kr
\item
  	Isaac Cho is with Hallym University (corresponding author).
  	E-mail: drisaaccho@gmail.com
}

\teaser{
  \centering
  \includegraphics[width=.95\linewidth]{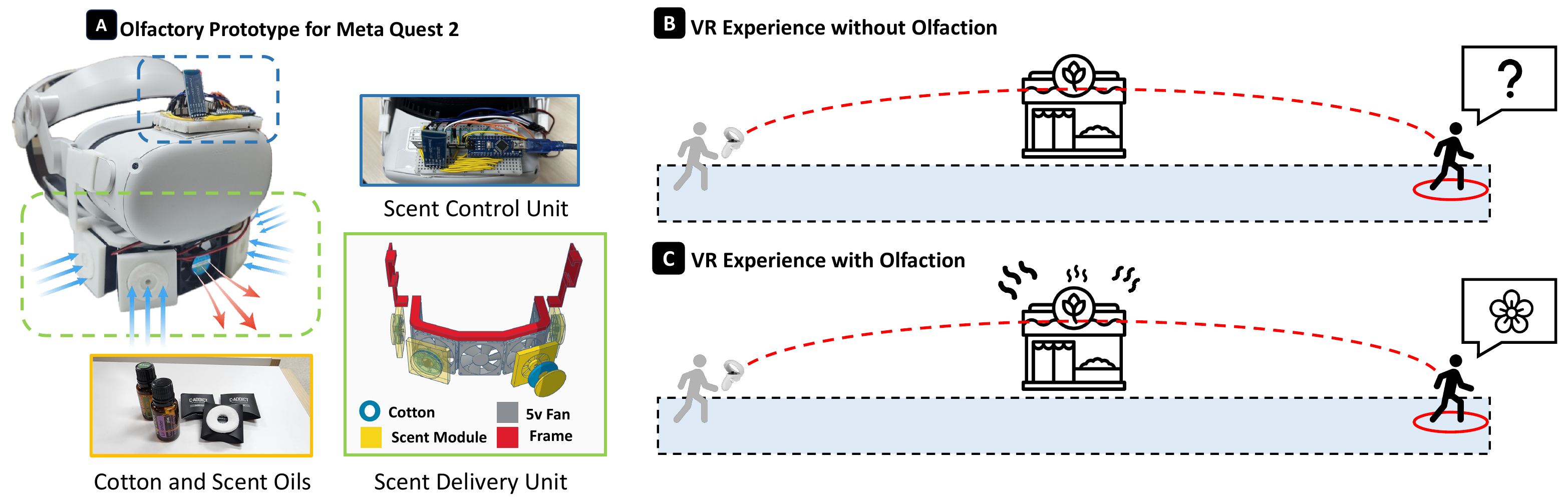}
  \caption{
  (A) The wearable olfactory device is capable of releasing distinct scents and incorporating a ventilation feature. 
  It consists of a Scent Control Unit and a Scent Delivery Unit.
   Scent Control Unit communicates with the VR system using an Arduino Nano and a Bluetooth module. 
Scent Delivery Unit has five fans, four of which are integrated with the scent module.
  Scent oils and cotton are used to store scents in the olfactory device.
 (B) During teleportation-based navigation in VR, users may have limited information about locations encountered during movement. Dash provides brief visual motion cues during navigation. (C) Olfactory cues can provide additional information that may help users recognize encountered locations during navigation.}
  \label{fig:teaser}
}

\abstract{
 Spatial cognition supports how people interpret spatial relationships and navigate their surroundings. 
 Although vision plays a dominant role, other sensory modalities, including olfaction, may also contribute under limited visual conditions. 
 This paper investigates the role of olfactory cues in location recognition during virtual reality (VR) navigation. 
 We conducted two formal studies using a wearable olfactory prototype. 
 Study 1 examined whether olfactory cues and ventilation influenced users' recognition of encountered locations during \HDY{teleportation-} and dash-based navigation. 
 Study 2 extended this investigation to repeated navigation and examined whether ventilation duration influenced recognition performance and user experience across successive movements. 
 Results show that olfactory cues can support recognition of encountered locations during navigation, while ventilation was associated with reduced residual interference and improved usability. 
 However, longer ventilation did not clearly improve recognition accuracy in repeated navigation. These findings suggest the potential of olfactory cues as contextual signals during VR navigation, while also highlighting the importance of scent management for maintaining perceptual clarity and user comfort.
} 


\begin{document}

\firstsection{Introduction}

\maketitle

Multisensory integration in Virtual Reality (VR) has been investigated as a means to enhance user immersion~\cite{fontaine1992experience, mcgreevy1992presence, witmer1998measuring}. 
Olfactory cues, consisting of chemical compounds in the air, interact with olfactory receptors in the human nose, enabling the perception of various scents~\cite{bushdid2014humans}. Detecting and perceiving olfactory cues is only a part of olfactory experiences. 
Previous research in cognitive science has shown that olfaction can trigger memories~\cite{chanes2016redefining, sullivan2015olfactory} and improve cognitive function~\cite{amores2017essence, barker2003improved, holland2005smells}. Furthermore, earlier studies have highlighted the potential benefits of integrating olfactory cues into VR~\cite{Bahremand2022SmellEngine, Jung2020IEEE, Fritzsch2020}.
For example, Sithu and Ishibashi~\cite{Sithu2017} and Nakamoto et al.~\cite{nakamoto2020virtual} reported that olfactory cues improve object recognition and increase immersion in VR. 
Navigation is a fundamental component of 3D user interaction and a common part of many VR experiences. Teleportation, a widely used navigation technique, allows users to instantly move to different locations in VR~\cite{al2018virtual}, making it particularly useful for traversing long distances beyond their physical space. 
However, due to its instantaneous movement, users' spatial awareness during virtual travel is diminished~\cite{prithul2021teleportation, Yao2023}.
To enhance spatial awareness during teleportation, various adaptations have been introduced, incorporating visual and auditory cues~\cite{bhandari2018teleportation, bolte2011jumper}. 
For example, Dash~\cite{bhandari2018teleportation, bolte2011jumper}, also known as the Jumper Metaphor, is a variation of teleportation that integrates visual motion cues. Instead of relocating the user instantaneously, it provides a smooth, rapid transition to the target location, which can improve spatial awareness and reduce disorientation during VR navigation.

Although these approaches can provide transitional visual or auditory information during locomotion, it remains unclear whether they sufficiently support users in recognizing or distinguishing virtual locations encountered during navigation.
This limitation may arise because VR technologies primarily rely on vision and audition, whereas human perception in the real world is multisensory. 
To the best of our knowledge, prior VR olfaction research has paid limited attention to teleportation-based locomotion, leaving it open how olfactory cues may be perceived and used during such navigation.
As an initial step toward addressing this research gap, this work investigates whether olfactory cues can be effectively perceived and utilized during teleportation-based navigation in VR.


In this work, we first introduce our olfactory device prototype, which is low-cost and attachable to a Meta Quest 2, capable of delivering up to four scents. The device also incorporates a ventilation fan to remove residual odors, addressing a challenge highlighted in prior work~\cite{Kato2019}.
Next, we present two user studies to evaluate the impact of integrating olfactory cues into teleportation-based techniques on the recognition of encountered locations during VR navigation.
In the studies, participants were asked to identify locations they had passed during single or multiple navigation trials while using teleportation or dash techniques, under conditions with and without olfactory cues. Our results suggest that incorporating olfactory cues can support participants' recognition of encountered locations during VR navigation. In addition, ventilation was helpful for scent management by reducing residual interference and improving usability in some conditions.
We conclude by discussing these findings, outlining study limitations, and proposing future directions for integrating olfactory cues with other 3D interaction techniques to create richer and more immersive VR experiences.
\section{Related Work}


\subsection{The Role of Olfaction}\label{sec_rw_role_olfaction}

The role of olfaction in daily life has been underestimated due to the belief that humans have a weaker sense of smell than other animals~\cite{schwarz2024human}.
However, recent research underscores the significant impact of olfaction on human spatial abilities. Dahmani et al.~\cite{dahmani2018intrinsic} provided neuroscientific evidence linking olfaction and spatial memory, while Jacobs et al.~\cite{jacobs2015olfactory} demonstrated that olfactory cues enhance participants' navigation in the real-world.

Previous research also showed that olfactory cues can evoke emotions~\cite{alaoui1997basic}, trigger memories~\cite{sullivan2015olfactory}, influence human behavior~\cite{ho2005olfactory, ilmberger2001influence}, and enhance cognitive function~\cite{amores2017essence, barker2003improved, holland2005smells}.
For instance, just as the smell of coffee activates odor representations in the brain, olfactory cues increase the salience of objects, making them easier to recognize~\cite{degel2001implicit}.
While most VR research has focused on visual and auditory stimuli to enhance user experience, there is growing interest in exploring multisensory interactions in VR~\cite{melo2020multisensory, fouad2025touching, bak2025beyond, kaur2025senses}. Understanding the role of olfactory cues in virtual environments is important for designing richer and more immersive experiences.

\subsection{Effectiveness of Olfactory Cues in VR}
%

Olfactory cues can enhance the VR user experience by introducing scents that evoke real-world associations, enriching immersion and engagement~\cite{melo2020multisensory, han2025if, bak2025beyond, narciso2020impact}.
Research has shown that olfactory stimuli can improve presence and immersion in VR, which is particularly beneficial for applications such as VR-based exposure therapy~\cite{narciso2020impact, weidner2023eating}.
Furthermore, previous studies suggest that the combination of artificial visual and olfactory stimuli can significantly influence human perception~\cite{han2025if}. For example, MetaCookie+\cite{narumi2011meta} introduced a device integrated with an AR-HMD, a tube, and a fan, which overlays a virtual cookie image onto a real cookie while presenting corresponding scents. This setup demonstrated that olfactory cues can alter the perceived taste of food. Similarly, Weidner et al.\cite{weidner2023eating} developed Smell-O-Spoon, investigating how a mismatch between visually perceived food and scents affects user perception.
Ranasinghe et al.~\cite{ranasinghe2020exploring} examined the use of peppermint scent in VR, demonstrating its effectiveness in reducing visually induced motion sickness.
Although numerous studies highlight the benefits of olfactory integration in VR, its role in 3D user interaction remains underexplored. 
Considering the role of olfaction in Sec.~\ref{sec_rw_role_olfaction}, this work explores whether providing olfactory cues during 3D user interaction, specifically teleportation-based navigation, can enhance users' recognition of encountered locations. More specifically, we investigate whether olfactory cues can support more accurate recognition of encountered locations despite the discrete and non-continuous nature of teleportation in VR.





\subsection{Olfactory Devices for VR}

Various olfactory devices have been introduced to enhance the VR experience by providing olfactory cues, making virtual environments more realistic and immersive~\cite{dinh1999evaluating, hoffman1998physically}. These devices can generally be categorized into stationary~\cite{heilig1962sensorama, yanagida2004projection, ischer2014incorporation, tsai2021does} and wearable types~\cite{niedenthal2019handheld, javerliat2022nebula, amores2017essence}.
Stationary olfactory devices are fixed in a specific location and provide users with scent stimuli within a designated area~\cite{heilig1962sensorama}. In contrast, wearable olfactory devices are designed to be worn on the body or integrated with VR or Augmented Reality (AR) Head-Mounted Displays (HMDs), enabling more personalized and dynamic scent delivery~\cite{patnaik2018information}. This section focuses on wearable olfactory devices.
Wearable olfactory devices deliver olfactory cues through fans or aerosolized fragrant liquids dispersed via air spraying~\cite{amores2017essence, brooks2020trigeminal, seah2014sensabubble}.
There are a few wearable type olfactory devices. For example, Brooks et al.~\cite{brooks2021stereo} proposed a device that attaches to the nose to simulate odors while warning the user of dangers such as burns or gas leaks. Niedenthal et al.~\cite{niedenthal2019handheld} introduced an attachable device for Vive controllers that provides users with scents for virtual objects as they interact. Recently, Javerliat et al.~\cite{javerliat2022nebula} introduced an affordable and open-source olfactory device called Nebula. It disperses scented air from cotton soaked in liquid using an Arduino-controlled fan. 
In this work, we introduce a prototype olfactory device, similar to Nebula, but equipped with more fans and capable of storing up to four scents, thereby providing richer olfactory cues in response to VR environments. 

\subsection{Teleportation in VR}


Navigation is a fundamental component of 3D user interactions.
Teleportation is widely used~\cite{bowman20043d} as it allows users to move instantly to the target position while remaining stationary in the real-world.
Although teleportation provides an efficient means of navigation in VR, Prithul et al.~\cite{prithul2021teleportation} and Cherep et al.~\cite{cherep2020individual} pointed out that users often experience difficulty immediately understanding their new spatial context after teleportation, which can lead to reduced immersion and spatial disorientation.

To mitigate this issue, previous research has augmented teleportation with transitional visual cues that help users maintain spatial context during navigation ~\cite{bolte2011jumper}.
They are known as dash approaches.
These approaches simulate a gradual transition to the destination rather than an instantaneous jump, allowing users to better perceive spatial relationships. 
Dash approaches have been explored using either dynamic speed adjustments~\cite{mackinlay1990rapid} or fixed-speed movement~\cite{bhandari2018teleportation}, enabling smoother and more natural transitions in VR. 
Bowman et al.~\cite{bowman1997travel} studied the impact of dash speed on users’ spatial understanding and compared dash-based navigation with teleportation.
They examined four different speed configurations including slow constant speed, fast constant speed, slow in/slow out, and immediate teleportation. 
Their findings demonstrated that immediate teleportation evoked the highest levels of disorientation and confusion, as the absence of transitional cues hindered users from effectively updating their spatial context. 
However, no significant differences were observed across the other speed conditions in terms of spatial awareness. Our work examines whether olfactory cues can improve recognition of encountered locations during VR navigation using teleportation and dash techniques.

\section{Olfactory Prototype for VR}\label{sec_wearableDevice}


This section introduces a prototype of a wearable olfactory device compatible with Meta Quest 2 (Fig.~\ref{fig:teaser})\footnote{Our Unity project and 3D print files for the prototype are available in our Git repository. \url{https://anonymous.4open.science/r/Wearable_Olfactory_Device-0B6C}}. 
It consists of upper and lower components, referred to as the \textit{Scent Control Unit} and \textit{Scent Delivery Unit}, which attach to the corresponding upper and lower sections of the HMD.
\textit{Scent Control Unit} manages scent emission and ventilation through communication with VR applications, while \textit{Scent Delivery Unit} houses the fans and positions the olfactory sources. 
This olfactory device can deliver up to four distinct scents, tailored to the device's ventilation function. 
It is designed with a focus on affordability (approximately \$50) and a form factor that can be attached to Meta Quest 2, weighing 250g.
Its dimension is 19.2cm (width) $\times$ 9cm (length) $\times$ 12.8cm (height).

The core components of \textit{Scent Control Unit} are an Arduino Nano 
and Bluetooth module (HC-06).
We selected Arduino Nano due to its compact form factor, lightweight, and low power consumption.
It is powered directly through Meta Quest 2 via a USB connection, eliminating the need for an external power source.
The Arduino Nano and Bluetooth module are mounted on a compact breadboard (8cm x 6cm). 
Communication between the device and the VR application for delivering and ventilating scents is handled via Bluetooth serial communication through the Arduino Nano.

\textit{Scent Delivery Unit} consists of five 5V fans, four scent modules, and supporting frames.
The scent modules and supporting frames are 3D-printed with PLA filament and assembled using adhesive (yellow and red parts in Fig.~\ref{fig:teaser}-A), while the fans are readily available commercial products.
Each fan measures 50mm $\times$ 50mm and operates at 5V. 
Four fans are designated for scent delivery, while the central fan functions exclusively as a ventilation unit. 
Installed in reverse orientation relative to the scent fans, the ventilation fan actively pulls air away from the user's face to remove lingering scents and circulate fresh air.
This ventilation feature was informed by prior work \cite{javerliat2022nebula}, which showed that actively extracting air can reduce residual odor, thereby preventing the next scent from mixing with the previous one. 
Cotton pads saturated with scented oils are placed inside the scent modules. 
When the device is activated, the fans aerosolize the scented oils, releasing fine odor particles into the airflow directed toward the user’s nose. 
By replacing cotton pads and scent cartridges, users can easily switch between different olfactory stimuli.



\begin{figure*}[t]
  \centering 
  \includegraphics[width=.95\textwidth]{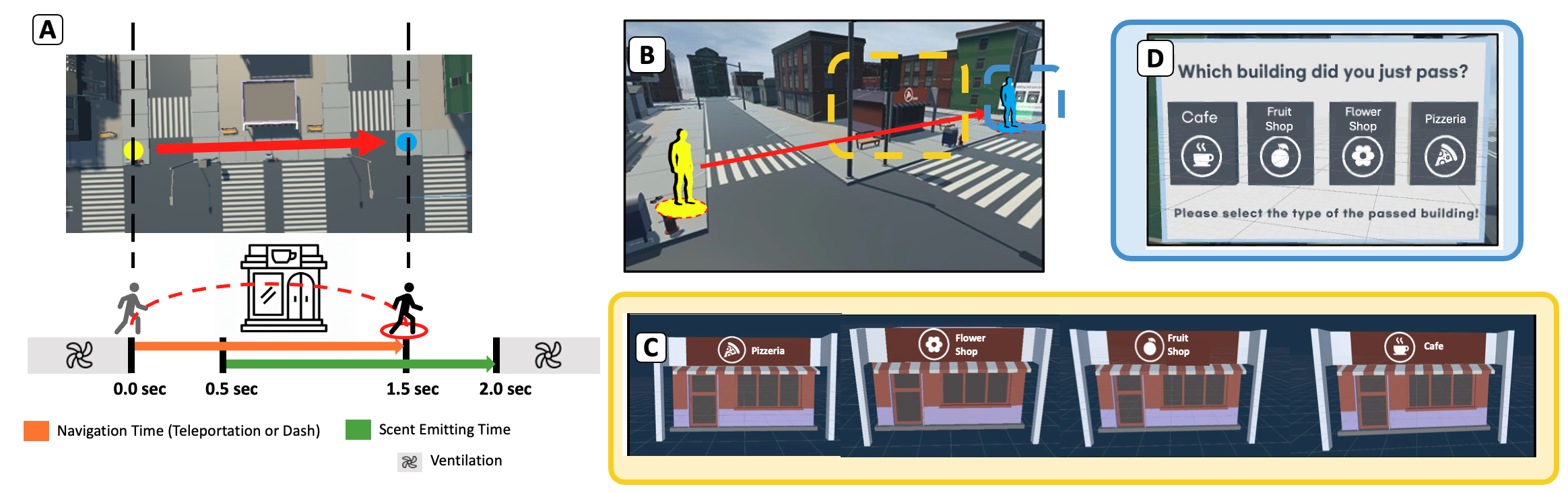}
  \vspace{-.2cm}
  \caption{
  (A) The navigating time from the starting point to the destination 
  is set to 1.5s. 
  After 0.5s of navigation activation, the scent fan fully operates, emitting a scent for 1.5s.
  Following that, the ventilation fan operates to remove the lingering scent until a new trial begins.
  (B) 
  A participant begins a task trial at the yellow mark and is asked to move to the blue mark using our conditions (i.e., TS+V, TS-V, DS+V, DS-V, and D). The participant passes by a store during the navigation. (C) The store is one of the following: a pizzeria, a flower shop, a fruit market, or a cafe. (D) After navigation, the participant is presented with a multiple-choice question and asked to choose the answer.  }
  \label{fig:Travel Experiment}
  \vspace{-.4cm}
\end{figure*}

Finally, the \textit{Scent Delivery Unit} connects to the \textit{Scent Control Unit} through jumper wires and MOSFET power transistors (IRFZ44N), which regulate fan activation and timing in synchronization with Unity events (Fig.~\ref{fig:Travel Experiment}-A). 
This modular separation between control and delivery simplifies maintenance. Once assembled, the left and right supporting-frame parts are securely attached to the head straps of the Meta Quest 2. 
To validate the olfactory prototype device before the main study, we conducted an informal pilot study with nine testers from our research lab.  \HDY{Additional information about the prototype and the preliminary pilot evaluation is available in~\cite{myung2023enhancing}.}



\section{Study 1: VR \HDY{Navigation} with Olfaction}


Study 1 \HDY{(\#IRB HIRB-2023-073).} examines the effects of integrating olfactory cues and ventilation features into teleportation-based navigation on recognition of encountered locations during a single navigation event in VR.
Our research question is as follows:
    \textit{``How do olfactory cues and ventilation influence recognition of encountered locations and user experience during instantaneous navigation in VR?''}


\subsection{Study Design}

We considered three factors: teleportation techniques, the presence of scents, and the presence of ventilation.
First, we considered two teleportation techniques: Teleportation and Dash.
Other navigation methods, such as walking and steering, were intentionally excluded because this study aims to isolate and evaluate issues specific to teleportation. 
These alternative methods also require continuous user input, which introduces variability in exposure to visual and olfactory stimuli and further complicates comparisons across conditions. 
Real-time control of olfaction for those continuous VR navigation techniques is challenging due to hardware limitations. 


\begin{itemize}[leftmargin=0.1in]
\item{\textbf{Teleportation}: It is widely used in VR applications because it allows users to navigate a virtual environment without their physical movement in the real-world.
When users activate teleportation by pointing to a target location on the floor with an input device, they are immediately transported to the target point~\cite{bowman1997travel}. 
In this work, Teleportation is accompanied by brief fade-out and fade-in visual effects.
However, Teleportation has been criticized for reducing immersion~\cite{clifton2020effects} and causing contextual losses due to its instantaneous movement~\cite{cherep2020individual}.
}

\item{\textbf{Dash}}:  
This technique is a variation of Teleportation that allows users to move rapidly to the desired location, rather than teleporting instantaneously~\cite{bhandari2018teleportation}. 
This enables users to visually perceive their movement~\cite{han2023evaluating, han2025perception}, which could improve engagement and spatial awareness, helping to mitigate the common criticisms of reduced immersion associated with Teleportation.
However, it could induce greater motion sickness due to the discrepancy between visual perception and physical sensations, known as sensory mismatch~\cite{weech2019presence}. 
In this work, a vignetting effect is applied to Dash to help mitigate motion sickness during travel by restricting Field of View (FoV) based on earlier research findings~\cite{bolas2017dynamic, lin2002effects, teixeira2021effects, fernandes2016combating}.
\end{itemize}





Considering the navigation techniques, presence of scents, and ventilation, we define the following five experimental conditions.
\begin{itemize}[leftmargin=0.1in, noitemsep]
    \item Teleportation with Scent and Ventilation (\textbf{TS+V})
    \item Teleportation with Scent but without Ventilation (\textbf{TS-V})
    \item Dash with Scent and Ventilation (\textbf{DS+V})
    \item Dash with Scent but without Ventilation (\textbf{DS-V})
    \item Dash without Scent and Ventilation (\textbf{D})
\end{itemize}



The \textit{D} condition served as the baseline for assessing the contribution of olfactory cues to store recognition during navigation. We did not include a teleportation-without-scent condition because, under instantaneous navigation, participants would have had little basis for store recognition beyond guessing. Instead, \textit{D} served as a practical baseline by preserving minimal visual information while keeping the recognition judgment interpretable.



\subsection{Task}~\label{sec_task}
We intentionally designed our task to be straightforward by limiting each trial to a single navigation event rather than multiple successive teleportations.
This decision was motivated by two key considerations. First, a simplified task structure enables us to isolate and better understand the specific effects of olfactory cues and ventilation on participants' recognition of encountered locations during VR navigation.
Second, incorporating multiple teleportations would likely require distinct scents for each navigation. 
This could introduce challenges related to scent layering, transition timing, and scent contamination. 
By controlling the teleportation to a single instance, we aim to preserve the clarity and focus of our study while minimizing the potential challenges.



For each task, participants navigate a VR scene from a starting point to a designated destination. Fig.~\ref{fig:Travel Experiment} illustrates the task. 
The scene depicts a city environment with multiple buildings and roads. 
The yellow and blue markers depicted in Fig.\ref{fig:Travel Experiment}-A and B indicate the starting and destination points, respectively, positioned 40 meters apart. 
A store is located between them, and it can be one of the following: a cafe, a fruit market, a flower shop, or a pizzeria (Fig.~\ref{fig:Travel Experiment}-C), all of which represent common, everyday locations. 
While the stores share identical structural designs, each possesses a unique scent and a clearly identifiable sign.
To provide unique scents, four distinct scents corresponding to the four stores are selected. 
They include a coffee scent for the cafe, a fruity scent for the fruit market, a floral scent for the flower shop, and a pizza scent for a pizzeria
~\footnote{All scent products were purchased from 
\url{https://www.serimfood.kr}.
}. 
At the starting point, participants can view only the side of the store, preventing them from identifying its type initially.
\HDY{Participants then initiate navigation by selecting a target position with the controller and travel from the starting point to the destination using either Teleportation or Dash.}
Depending on the task condition, they may or may not experience scents while traveling to the destination, depending on the navigation condition. During Dash, participants could potentially glimpse the store briefly despite the rapid movement and vignette effect.
Upon reaching the destination, participants are presented with a multiple-choice question: \textit{``Which building did you just pass?''} featuring four options, as illustrated in Fig.~\ref{fig:Travel Experiment}D. 
After answering the question, participants either proceed to the next task or complete the study.

An overview of navigation and olfactory device operation is depicted in Fig.~\ref{fig:Travel Experiment}-A. 
We carefully set the navigation and scent dispersion duration to 1.5 s to ensure effective scent delivery. 
\HDY{This duration was selected for two reasons. 
First, participants need sufficient exposure time to recognize a scent. 
A natural human sniff typically lasts approximately 1.6 s~\cite{laing1983natural}, and prior research has shown that individuals require approximately 1 s to reliably detect and identify a scent~\cite{olofsson2014time}.
}
Second, the study by Rahimi et al.~\cite{rahimi2018scene} supports our design choice. They compared instant teleportation with teleportation incorporating a 1.5 s fade-in and fade-out effect and found no significant difference in VR sickness or user preference. This finding supports the validity of our use of a 1.5 s teleportation duration.
Additionally, we observed that the device’s fans required approximately 0.5 s to fully activate. To ensure consistent delivery of visual and olfactory stimuli for the same duration, we extended fan operation by an additional 0.5 s after navigation ended, totaling 1.5 s of scent emission.
In conditions where ventilation is used, the ventilation fan activates immediately after scent emission concludes and remains on continuously until the start of the next trial.

Please note that the scent delivery and ventilation fans were not intentionally controlled to adjust diffusion levels or scent intensity by modifying their operation cycles. The fans operated in a simple binary mode (either `on' or `off') without intermediate settings.

\subsection{Measurement and Hypothesis}
One \HDY{objective} and three subjective measurements are employed.
The \HDY{objective} metric used is the accuracy of participants' responses to the questions. 
The three subjective metrics include System Usability Scale (SUS), NASA Task Load Index (NASA-TLX), and Simulator Sickness Questionnaire (SSQ).

\begin{itemize}[leftmargin=0.1in, noitemsep]


    \item \textbf{Task Accuracy:} This quantitative metric represents the proportion of correct responses across 12 trials per condition, \HDY{normalized to a scale ranging from 0 (all responses incorrect) to 1 (all responses correct).}.


    \item \textbf{SUS:} This questionnaire assesses the usability of the technology. It includes 10 questions with 5 response options each (1: Strongly Disagree - 5: Strongly Agree), which collectively sum to a maximum score of 100 points~\cite{brooke1996sus}.
    

    \item \textbf{NASA-TLX:} Participants’ subjective workload was assessed using four subscales from the NASA Task Load Index (NASA-TLX) questionnaire~\cite{hart1988development}: Mental Demand, Physical Demand, Sense of Frustration, and Performance. 
    Mental Demand reflects perceived cognitive load, while Physical Demand captures physical workload during navigation.
    Sense of Frustration reflects feelings of difficulty or discomfort when navigating with scents. 
    Performance represents perceived task success.
    Each is assessed on a 20-point Likert scale\Note{, with lower scores indicating lower workload or better performance. }

    \item \textbf{SSQ:} It measures the degree of VR sickness (i.e., nausea, disorientation, and oculomotor discomfort), through 16 questions~\cite{kennedy1993simulator}. In our study, we used a 5-point Likert scale to allow users to choose a neutral score.
    The increase in VR sickness was assessed by computing the SSQ score difference between the baseline and post–sub-session measurements for each of the five conditions.
    
\end{itemize}



    

\noindent
For the measurements, we established the following hypotheses to answer our research question:

\begin{enumerate}[label=H\arabic*., noitemsep]

 \item Incorporating olfactory cues during navigation is expected to improve recognition accuracy for encountered locations by providing additional sensory cues.
    \item Teleport and Dash with both scent and ventilation (TS+V and DS+V) would exhibit higher accuracy and greater user satisfaction, as reflected in higher SUS scores and improved NASA-TLX, compared to these with scent only and no ventilation (TS-V and DS-V). This improvement is attributed to ventilation effectively preventing scent mixing, leading to clearer sensory differentiation and enhanced outcomes.
   

  \item Providing both olfactory and visual cues (DS+V and DS-V) is expected to achieve higher task accuracy compared to Teleportation with Scent conditions (TS+V and TS-V), where only olfactory cues are available.

    \item Incorporating olfactory cues into VR navigation would reduce motion sickness, as evidenced by lower SSQ scores~\cite{ranasinghe2020exploring}.





  
\end{enumerate}

\subsection{Participants}

We initially recruited a total of 24 participants. 
However, four participants were excluded from the analysis: one withdrew due to severe dizziness during the task, and three provided insufficient responses to the questionnaires (e.g., assigning a uniform score of zero to all items). 
 As a result, records from 20 participants were analyzed (11 males, 9 females, age M=23.17, ranging from 21 to 25, SD=0.76). 
They had normal or corrected-to-normal vision (20/20) and reported no impairments affecting their use of VR devices or olfactory perception.
The participants were compensated about \$7.50 for their participation, which reflected the local hourly minimum wage in the country where the study was conducted.

%

\subsection{Apparatus and Study Space}
Meta Quest 2 and the introduced wearable olfactory device are used. Meta Quest 2 has a 104$^\circ$ horizontal and 98$^\circ$ vertical FoV, and a resolution of 1832 \texttimes{} 1920 pixels per eye.  
The virtual scene for the study was developed in Unity 3D \HDY{(2021.3.17.f1)} and run on a Windows 11 desktop equipped with an AMD Ryzen 5 4600H CPU, 16GB RAM, and a GeForce RTX 2060.  The application ran on the Quest 2 via AirLink. 
The study was conducted in a room measuring 4.3m $\times$ 2.9m $\times$ 3m with windows. The windows were only opened during break periods between sub-sessions.
The olfactory device remained mounted on the HMD throughout the experiment for all conditions.



\subsection{Procedures}

The study is a 60-minute-long session including 1) reading and signing a consent form, 2) completing a demographic questionnaire, 3) participating in a training session with instructional guidelines, and 4) performing the main session. All sessions are conducted with participants seated in a chair. 

Upon arrival of a participant, the participant signs the informed consent form.
The participant then completes a demographic questionnaire and receives a description of the study procedures and tasks, along with an introduction to the VR HMD and olfactory device. The questionnaire inquires about gender, age, and whether the participant has any issues with their sense of smell. The participant also completes the initial SSQ questionnaire to assess their baseline condition. 
Thereafter, the participant undergoes a training session to learn the navigation techniques, preview the building signage and scents used in the study, and receive instructions on using the controller to answer multiple-choice questions in VR.
The training session takes about 5 minutes, and is followed by 5 minutes of freshening the study room by opening windows before the main session.

The main session includes 5 sub-sessions. In each sub-session, the participant performs the task 12 times (4 scents \texttimes{} 3 trials each) using one of the 5 navigation conditions. 
As a result, the participant completes a total of 60 trials during the main session. 
The order of the navigation techniques is fully counterbalanced, while scents are presented in a random sequence. 
After completing each sub-session, the participant answers the SUS, NASA-TLX, and SSQ questionnaires \HDY{based on their overall experience performing the task.} 
Before proceeding to the next sub-session, participants take a 5-minute break to recover their sense of smell.
During the break, we refresh the study room by opening the windows and using an air circulator to clear any lingering scents from the previous sub-session. Additionally, the ventilation fan is operated to remove any residual scents from the device. This helps minimize cross-condition scent contamination and sensory fatigue.



\begin{figure}[t]
   \centering 
   \includegraphics[width=.9\columnwidth]
   {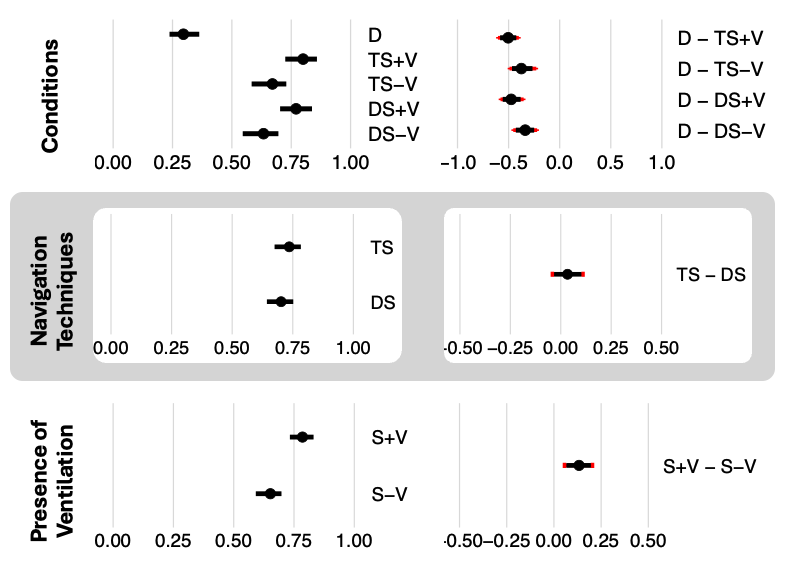}
   \vspace{-.2cm}
   \caption{Study 1 - Task Accuracy Results. 
   Each row shows average results first and pairwise comparison results next, with error bars. 
   The x-axis indicates normalized accuracy, ranging from 0 (all incorrect) to 1 (all correct). The error bars represent 95\% Bootstrap confidence intervals (CIs). 
   Adjusted CIs for the pairwise comparisons with Bonferroni correction are shown in red. 
   }
   \label{Result Graph}
  \vspace{-.4cm}
 \end{figure}

\subsection{Results}

\HDY{To analyze the effects of the study conditions, we adopted an estimation-based approach by reporting the mean values and their 95\% confidence intervals (CIs)~\cite{calmettes2012making, dragicevic2016fair}.}
The 95\% CIs were computed using the bias-corrected and accelerated (BCa) bootstrap method with 10,000 resamples.
\HDY{The bootstrap method was chosen because it does not rely on strong distributional assumptions, providing a robust approach for estimating confidence intervals.}

\begin{figure}[t]
   \centering 
   \includegraphics[width=.9\columnwidth]{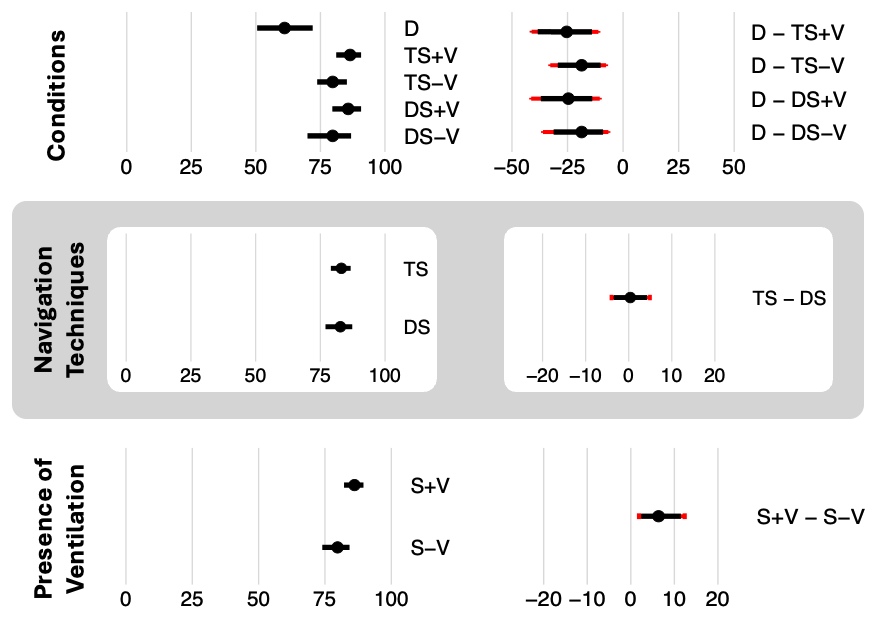}
   \vspace{-.2cm}
   \caption{Study 1 - SUS Results with 95\% CI error bars. Each row shows the measurement average first and pairwise comparison results next.
   }
   \label{fig:SUS Graph}
   \vspace{-.4cm}
 \end{figure}

\HDY{To evaluate differences between conditions, we report the mean differences and their Bonferroni-adjusted 95\% CIs~\cite{higgins2004introduction}.
In the result figures (i.e., Fig.~\ref{Result Graph}-\ref{fig:ssq_result}), adjusted CIs are shown as red error bars, whereas unadjusted CIs are shown as black error bars. 
Mean differences whose adjusted CIs do not include 0 are considered statistically significant.
Beyond statistical significance, CIs offer greater interpretive depth by conveying both the magnitude of mean differences and the precision of the estimates~\cite{cockburn2020threats,dragicevic2016fair}.
A narrow CI indicates a more precise estimate, whereas a wide CI indicates greater uncertainty about the magnitude and direction of the effect.
}



We first report the mean and 95\% CI values for the conditions, along with pairwise comparisons against the baseline (i.e., D).
We also group the conditions by the technique types with scents (TS and DS) or the presence of ventilation (S+V and S-V) and evaluate their effects in the same manner. 
\HDY{Detailed baseline comparison and group analysis results are provided in the supplementary material (Tables~\ref{study1_all_results_baseline_comparison} and~\ref{study1_all_results_group_analysis}).}



\subsubsection{Task Accuracy}
 Fig~\ref{Result Graph} presents the results. Accuracy was highest in TS+V (0.80, \HDY{\ciValues{CI=[0.72, 0.86]}}), followed by TS-V (0.67, \HDY{\ciValues{CI=[0.58, 0.73]}}), DS+V (0.77, \HDY{\ciValues{CI=[0.70, 0.84]}}), DS-V (0.63, \HDY{\ciValues{CI=[0.55, 0.70]}}), and D (0.29, \HDY{\ciValues{CI=[0.24, 0.36]}}). 
Pairwise comparisons revealed clear evidence supporting \textbf{H1}: D had a lower accuracy than TS+V by 0.50, \HDY{\ciValues{CI=[0.40, 0.60]}}, TS-V by 0.38, \HDY{\ciValues{CI=[0.23, 0.49]}}, DS+V by 0.47, \HDY{\ciValues{CI=[0.35, 0.58]}}, and DS-V by 0.33, \HDY{\ciValues{CI=[0.22, 0.45]}}. 

Next, group analysis results (Fig~\ref{Result Graph}) further showed that  S+V (0.79, \HDY{\ciValues{CI=[0.73, 0.83]}}) has a clearly higher accuracy than S-V (0.65, \HDY{\ciValues{CI=[0.59, 0.70]}}) by 0.13, \HDY{\ciValues{CI=[0.06, 0.20]}}, supporting our \textbf{H2}. However, TS (0.74, \HDY{\ciValues{CI=[0.68, 0.78]}}) and DS (0.70, \HDY{\ciValues{CI=[0.64, 0.75]}}) had no clear difference (0.03, \HDY{\ciValues{CI=[-0.04, 0.11]}}), rejecting \textbf{H3}.


\subsubsection{\HDY{System Usability Scale} (SUS)}
The results are reported in Fig.~\ref{fig:SUS Graph}: TS+V (86.5, \HDY{\ciValues{CI=[81.12, 90.75]}}), TS-V (79.75, \HDY{\ciValues{CI=[73.75, 85.25]}}), DS+V (85.75, \HDY{\ciValues{CI=[79.62, 90.75]}}), DS-V (79.75, \HDY{\ciValues{CI=[70.00, 86.87]}}), and D (61.12, \HDY{\ciValues{CI=[50.50, 72.00]})}.
Pairwise comparisons showed that D had a clearly lower SUS than TS+V by 25.37, \HDY{\ciValues{CI=[11.12, 42.25]}}, TS-V by 18.62, \HDY{\ciValues{CI=[7.65, 32.75]}}, DS+V by 24.62, \HDY{\ciValues{CI=[10.50, 41.34]}}, and DS-V by 18.62, \HDY{\ciValues{CI=[6.62, 36.03]}}, supporting \textbf{H2}.

The group analysis showed that S+V (86.13, \HDY{\ciValues{CI=[82.19, 89.5]}}) is 6.38, \HDY{\ciValues{CI=[1.88, 12.38]}} higher than S-V (79.75, \HDY{\ciValues{CI=[74.00, 84.25]}}). However, TS (83.12, \HDY{\ciValues{CI=[79.06, 86.69]}}) and DS (82.75, \HDY{\ciValues{CI=[77.00, 87.31]}}) showed no clear difference (0.38, \HDY{\ciValues{CI=[-4.00, 4.94]}}).


\begin{figure}[t]
   \centering 
   \includegraphics[width=.9\columnwidth]{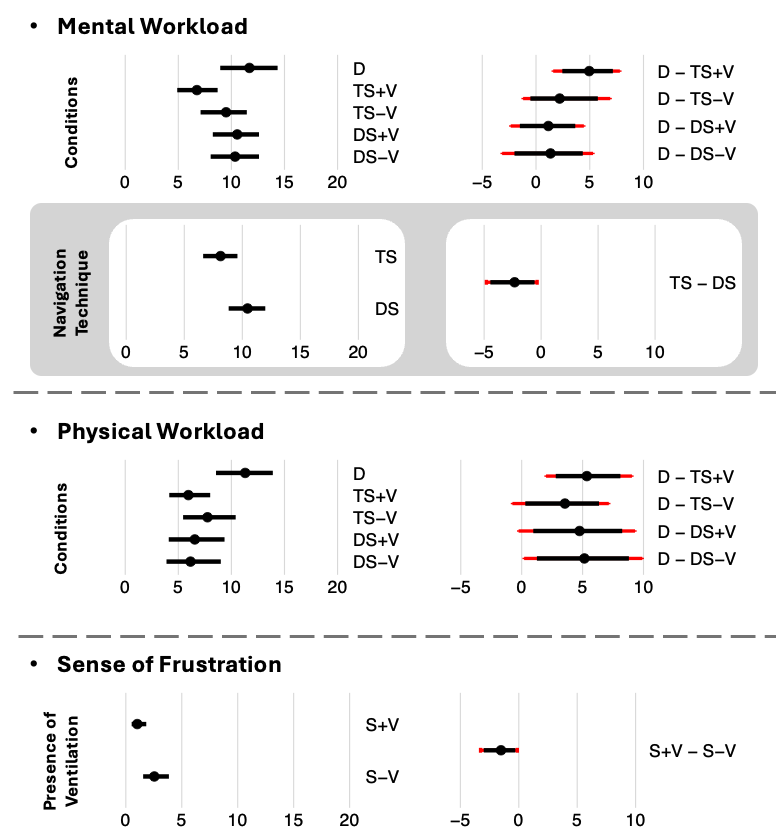}
   \vspace{-.2cm}
   \caption{Study 1 - NASA-TLX Results (Mental Workload, Physical Workload, and Sense of Frustration) with 95\% CI error bars. 
   Each row shows the measurement average first and the pairwise comparison results next. 
   }
   \label{fig:NASA-TLX Graph}
   \vspace{-.4cm}
 \end{figure}

\subsubsection{NASA-TLX}

The NASA-TLX results are shown in Fig.~\ref{fig:NASA-TLX Graph}. 
 
\textbf{Mental Workload}
No clear difference between the conditions was disclosed, showing  TS+V:6.75, \HDY{\ciValues{CI=[4.90, 8.70]}}, TS-V:9.50, \HDY{\ciValues{CI=[7.10, 11.45]}}, DS+V:10.55, \HDY{\ciValues{CI=[8.25, 12.60]}}, DS-V:10.35, \ciValues{CI=[8.05, 12.60]}, and D:11.70, \HDY{\ciValues{CI=[8.95, 14.35]}}.

The group analysis results were TS:8.13, \HDY{\ciValues{CI=[6.63, 9.58]}}, DS:10.45, \HDY{\ciValues{CI=[8.83, 11.98]}}, S+V:8.65, \HDY{\ciValues{CI=[7.18, 10.20]}}, and S-V:9.93, \HDY{\ciValues{CI=[8.28, 11.48]}}.
TS showed clearly lower mental load than DS by 2.33, \HDY{\ciValues{CI=[0.35, 4.80]}}. However, no clear difference between S+V and S-V was found (1.275, \HDY{\ciValues{CI=[-3.45, 1.13]}}).




 
\textbf{Physical Workload}
The results were TS+V:5.95, \HDY{\ciValues{CI=[4.15, 8.00]}}, TS-V:7.75, \HDY{\ciValues{CI=[5.45, 10.40]}}, DS+V:6.55, \HDY{\ciValues{CI=[4.10, 9.35]}}, DS-V:6.15, \HDY{\ciValues{CI=[3.89, 9.00]}}, and D=11.3, \HDY{\ciValues{CI=[8.55, 13.90]}}. Their pairwise comparison results showed that TS+V was 5.35, \HDY{\ciValues{CI=[2.00, 9.05]}} lower than D, while DS-V was 5.15, \HDY{\ciValues{CI=[0.20, 9.90]}} lower than D.

However, the group analysis revealed no clear differences: TS:6.85, \HDY{\ciValues{CI=[5.38, 8.50]}}, DS:6.35, \HDY{\ciValues{CI=[4.68, 8.28]}}, S+V:6.95, \HDY{\ciValues{CI=[4.73, 7.90]}}, and S-V:6.95, \HDY{\ciValues{CI=[5.30, 8.83]}.}




\textbf{Sense of Frustration} 
We found no differences across the conditions: TS+V:0.70, \HDY{\ciValues{CI=[0.20, 1.55]}}, TS+V:2.70, \HDY{\ciValues{CI=[1.20, 5.10]}}, DS+V:1.35, \HDY{\ciValues{CI=[0.60, 2.85]}}, DS-V:2.40, \HDY{\ciValues{CI=[1.30, 4.00]}}, and D:2.70, \HDY{\ciValues{CI=[1.25, 5.01]}}.

In the group analysis, while no clear difference was found between the navigation groups (TS:1.75, \HDY{\ciValues{CI=[0.88, 3.00]}} and DS:1.88, \HDY{\ciValues{CI=[1.18, 2.95]}}), a significant difference was observed based on the presence of ventilation. S+V (1.03, \HDY{\ciValues{CI=[0.53, 1.83]}}) was notably lower than S-V (2.55, \HDY{\ciValues{CI=[1.55, 3.85]})} by 1.53, \HDY{\ciValues{CI=[0.13, 3.28]}}.



\textbf{Performance} 
\Note{The results were TS+V:4.80, \HDY{\ciValues{CI=[3.10, 6.55]}}, TS-V:9.45, \HDY{\ciValues{CI=[7.25, 11.70]}}, DS+V:7.70, \HDY{\ciValues{CI=[5.60, 9.65]}}, DS-V:7.25, \HDY{\ciValues{CI=[5.45, 8.65]}}, and D=14.40, \HDY{\ciValues{CI=[11.45, 16.55]}}. 
Their pairwise comparison showed that all scented conditions were lower than D. }
\Note{However, we found no difference in the group analysis. }

\textbf{In summary,} we found minimal improvements from providing olfactory cues and ventilation features. However, no significant differences were observed between S+V and S-V, leading to the rejection of \textbf{H2} in terms of NASA-TLX.

\subsubsection{\HDY{Simulator Sickness Questionnaire} (SSQ)}


\textbf{Nausea} Participants reported increased nausea after TS-V (0.9, \HDY{\ciValues{CI=[0.05, 1.80]}}), DS-V (1.05, \HDY{\ciValues{CI=[0.35, 1.80]}}), and D (0.95, \HDY{\ciValues{CI=[0.20, 2.00]}}), while no clear differences were observed for TS+V (0.50, \HDY{\ciValues{CI=[-0.45, 1.80]}}) and DS+V (-0.30, \HDY{\ciValues{CI=[-2.55, 0.60]}}). However, we found no clear differences in the magnitude of these changes.



\begin{figure}[t]
  \centering 
  \includegraphics[width=.9\linewidth]{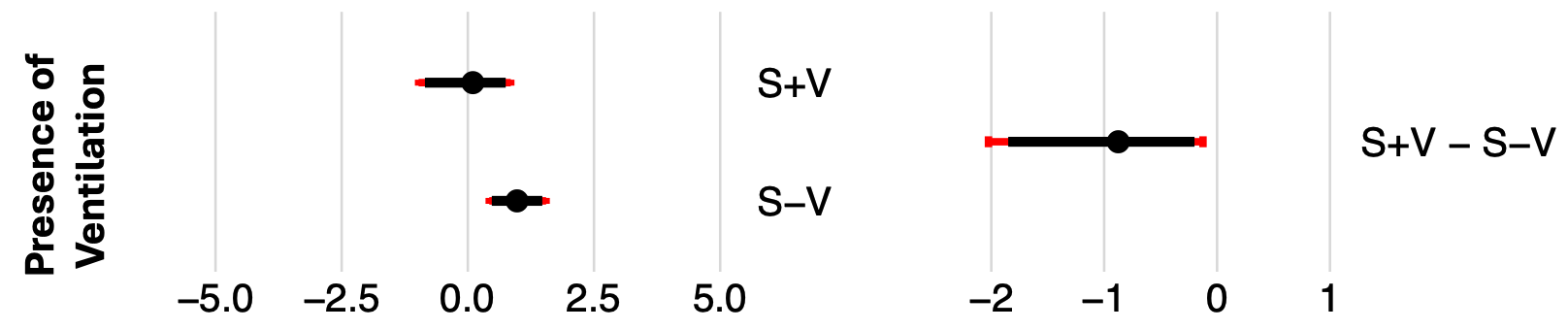}
  \caption{Study 1 - SSQ Nausea Results with 95\% CI error bars.
  It shows the measurement average first and pairwise comparison results next.
  The average results represent the difference between the post- and baseline-measurement. 
  }
  \label{fig:ssq_result}
  \vspace{-.4cm}
\end{figure}

In the group analysis, we found increases after TS (0.70, \HDY{\ciValues{CI=[0.03, 1.43]}}) and S-V (0.98, \HDY{\ciValues{CI=[0.43, 1.55]}}), while no differences was observed for DS (0.38, \HDY{\ciValues{CI=[-0.75, 0.98]}}) and S+V (0.10, \HDY{\ciValues{CI=[-0.98, 0.85]}}).
Please note that S+V yielded clearly smaller changes than S-V by 0.875, \HDY{\ciValues{CI=[0.125, 2.025]}}. The results are shown in  Fig.~\ref{fig:ssq_result}.

\textbf{Disorientation} We found clear increase after DS-V (1.05, \HDY{\ciValues{CI=[0.45, 1.85]}}) and D (0.85, \HDY{\ciValues{CI=[0.10, 1.75]}}), while the other conditions showed no significant differences (DS+V: -0.10, \HDY{\ciValues{CI=[-1.50, 1.00]}}, TS+V: 0.40, \HDY{\ciValues{CI=[-0.50, 2.20]}}, and TS-V: 0.20, \HDY{\ciValues{CI=[-0.30, 0.70]}}).
However, we found no clear differences in the magnitude of these changes.

The group analysis results revealed an increase of 0.63, \HDY{\ciValues{CI=[0.20, 1.15]}} in S-V, while no clear differences were observed for the other groups (S+V: 0.15, \HDY{\ciValues{CI=[-0.68, 1.1]}}, TS: 0.3, \HDY{\ciValues{CI=[-0.23, 1.21]}}, and DS: 0.48, \HDY{\ciValues{CI=[-0.35, 1.13]}}). However, no significant difference in the amount of changes was reported.

\textbf{Oculomotor Discomfort} 
Participants showed no clear difference after using DS+V (0.65, \HDY{\ciValues{CI=[-0.85, 2.15]}}), while the other conditions showed notable increases. TS+V was increased by 1.75, \HDY{\ciValues{CI=[0.25, 3.80]}}, TS-V by 1.45, \HDY{\ciValues{CI=[0.40, 2.80]}}, DS-V by 1.80, \HDY{\ciValues{CI=[0.25, 3.65]}}, and D (1.55, \HDY{\ciValues{CI=[0.10, 3.50]}}). However, the magnitude of these changes were comparable and no clear differences were observed.

Next, both TS and DS showed clear increases of 1.60, \HDY{\ciValues{CI=[0.65, 2.80]}} and 1.23, \HDY{\ciValues{CI=[0.13, 2.44]}}, respectively. Similarly, S+V and S-V increased by 1.20, \HDY{\ciValues{CI=[0.13, 2.48]}} and 1.63, \HDY{\ciValues{CI=[0.65, 2.78]}}, respectively. However, no difference in the amount of changes was reported.

\textbf{In summary,} we found some evidence of clear increments after performing tasks across the conditions. 
However, the magnitude of increase did not differ significantly between the conditions. This leads to the rejection of \textbf{H4}.

\subsection{Study 1: Discussions}




\textbf{Olfactory cues supported recognition of encountered locations during teleportation-based navigation in VR, but adding visual motion cues did not yield a clear accuracy benefit.}
Participants showed higher recognition accuracy in all scented conditions than in \textit{Dash}.
However, we found no significant accuracy difference between Teleport-S and Dash-S.
This result was unexpected, as we expected that having more sensory cues (i.e., visual and olfactory cues together in this work) would result in higher accuracy.

One possible explanation is that combining rapid visual input with transient olfactory cues increased attentional demands during \HDY{navigation~\cite{bhandari2018teleportation}.}
In this context, Dash may have increased cognitive load by requiring participants to process rapid visual transitions alongside transient olfactory signals, which may have contributed to lower task accuracy.
Our NASA-TLX results and participant feedback support this interpretation. 
Study 1 Participants reported significantly higher mental workload in DS than in TS, consistent with the findings of Bhandari et al.~\cite{bhandari2018teleportation}. 
They also noted that Dash felt less comfortable due to the need to process visual and olfactory cues simultaneously.

We also found \textit{Dash} resulted in higher average physical workload compared to the other scented navigation conditions. Unlike the scented conditions, where participants could partially rely on olfactory cues to aid target recognition, those in Dash had to rely solely on visual information. 
Given the fixed navigation duration of 1.5 seconds in this study, Dash involved a travel with the velocity of approximately 26.7m/s (= 40m / 1.5s).
We observed that participants frequently rotated their heads in an attempt to identify visual targets during the Dash navigation.
This increased reliance may have led to more frequent head movements, contributing to greater perceived physical effort.
\textbf{Adding ventilation was associated with higher recognition accuracy and better usability ratings than scent-only conditions.}
Providing olfactory cues with ventilation (Teleport-SV and Dash-SV) was associated with higher task accuracy, better usability, and lower frustration than the corresponding scent-only conditions (Teleport-S and Dash-S).
This pattern may reflect the role of ventilation in reducing lingering scents between transitions, which may have helped preserve perceptual separation between olfactory cues.
This finding is well aligned with the results of Javerliat et al.~\cite{javerliat2022nebula} and Maggioni et al.~\cite{Maggioni2020}, highlighting the importance of ventilation in eliminating lingering scents.

\textbf{Providing olfactory cues does not always alleviate motion sickness.}
Previous research~\cite{ranasinghe2020exploring} reported that olfactory cues, particularly peppermint scent, have a positive effect in reducing motion sickness. 
However, our SSQ results in Study 1 do not support that incorporating olfactory cues alleviates motion sickness in either Teleportation or Dash.
This discrepancy may be attributed to the choice of scents. 
On one hand, certain scents like peppermint~\cite{ranasinghe2020exploring, moss2008modulation}, are known to possess refreshing or soothing properties, whereas others may not have the same effect. 
On the other hand, some strong scents may even aggravate symptoms for certain individuals~\cite{bell1996neuropsychiatric}.

\section{Study 2: Repeated Teleportation with Olfaction}

Study 2 \ic{(IRB \#HIRB-2026-002)} extends Study 1 from a single-navigation setting to repeated navigation, where users encounter multiple successive movements within a trial. This extension is important because residual scent and perceptual interference are more likely to accumulate across repeated navigations than in a single movement. Since Study 1 suggested that ventilation may help preserve perceptual separation between scents and improve some user experience measures, Study 2 examines whether similar benefits emerge in repeated navigation and whether ventilation duration further influences recognition performance and user experience.
More specifically, our research question in Study 2 is:
    \textit{``Can the benefits observed in single teleportation be extended to repeated navigation, and how does ventilation duration influence recognition performance for encountered locations across successive teleportations?''}


\subsection{Study Design and Task}


Study 2 adopts a within-subjects design and uses the same navigation techniques (i.e., \textbf{Teleportation} and \textbf{Dash}) as independent variables. \ic{Unlike Study 1, which evaluated a single navigation event, Study 2 evaluates participants' recall of sequential navigation experiences across repeated navigation events. }

\begin{figure}[t]
  \centering 
  \includegraphics[width=\linewidth]{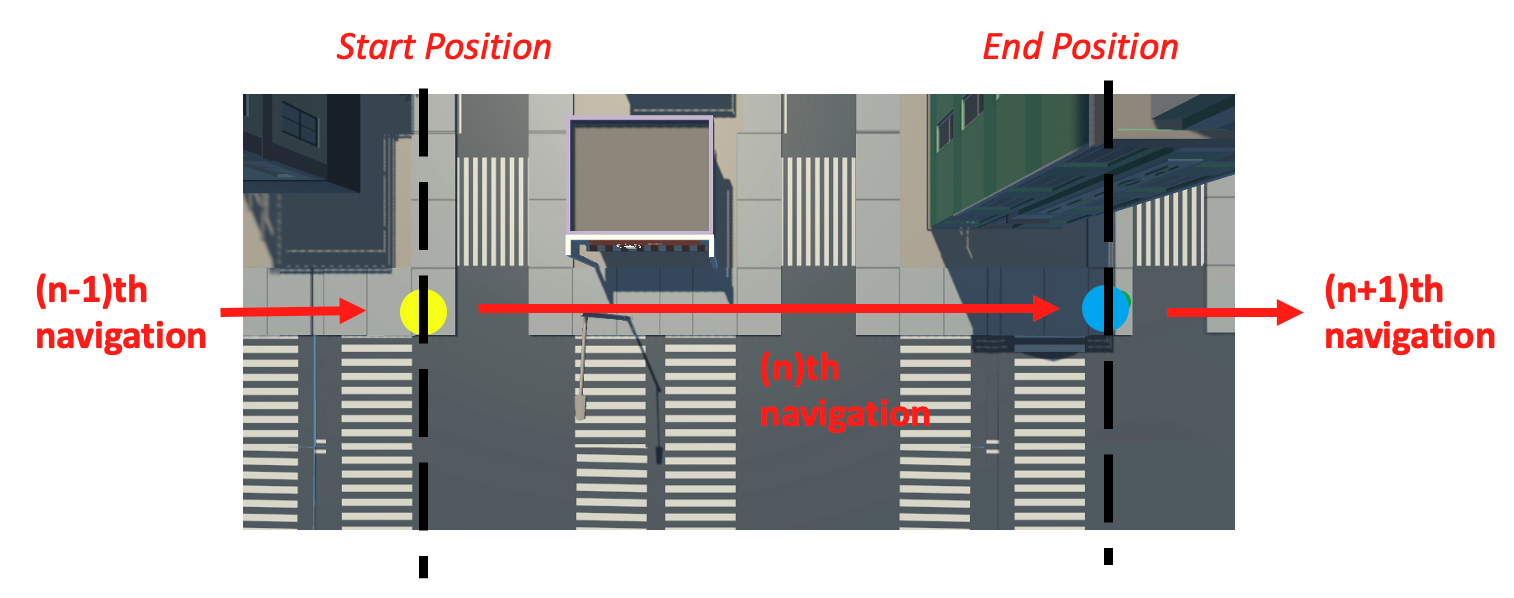}
  \vspace{-.5cm}
  \caption{\HDY{Navigation path configuration in Study 2. The same VR buildings from Study 1 were used, but the scene layout and navigation path were modified. The navigation endpoint was shifted while the start position remained fixed, and the city blocks were proportionally scaled down to preserve the navigation speed while ensuring the same sequence of visual scenes.}
}
  \label{fig:study2_scene}\vspace{-.4cm}
\end{figure}

To address our research question, Study 2 introduces two additional independent variables: \textit{the number of navigations} and \textit{ventilation duration}. \ic{The number of navigations corresponds to four successive movements within each trial: the first movement (\textbf{MV1}), second movement (\textbf{MV2}), third movement (\textbf{MV3}), and fourth movement (\textbf{MV4}).}  Although the number of navigations is not treated as an explicit experimental condition. It serves as an independent variable to examine how accurately participants remember locations encountered across successive navigations. Ventilation duration was controlled at two levels: 1 and 2 seconds (\textbf{Vent1S} and \textbf{Vent2S}). 
The ventilation fan was activated for the specified duration, starting at the time indicated in Fig.~\ref{fig:Travel Experiment}-A. 
To avoid potential effects caused by time variations between subsequent navigations and to account for fan activation timing, participants remained in the same position for 2.5s after completing one navigation before beginning the next. 
As a result, the interval between successive navigations was kept constant, while the duration of fan activation differed between conditions.

\HDY{
As in Study 1, the same VR assets and scent types were used. However, Study 2 differed from Study 1 in two respects. First, the number of navigations per trial was increased from one to four (i.e., MV1, MV2, MV3, and MV4). Second, the navigation end positions and scene layout were adjusted to ensure that participants observed the same sequence of visual scenes during each navigation.
Specifically, the navigation end position was shifted while the start position remained unchanged as shown in Fig.~\ref{fig:study2_scene}.
To maintain the same navigation speed as in Study 1, we proportionally reduced the size of each city block so that the shifted navigation end position did not change the overall navigation distance.
As a result, the navigation speed remained identical to that in Study 1, while ensuring that participants traversed the same visual sequence during each navigation.
%
}
Aside from these two additions, all other settings, including navigation duration and scent emission, remained identical to those in Study 1.

In each trial of Study 2, participants performed four consecutive navigations using one of the two navigation techniques. 
\HDY{
Once participants initiated the first movement using the controller, the subsequent movements were triggered automatically to ensure consistent ventilation durations (i.e., Vent1S and Vent2S) between movements.}
Each movement was accompanied by scent emission followed by a designated ventilation period. 
Across the movements, scent repetition was allowed, but the same scent was not presented in consecutive movements.
After the four movements, participants were asked to select the buildings they had passed in the order they encountered them.
The trial then concluded with a final ventilation phase to eliminate any residual scent and prepare the environment for the subsequent trial.

\subsection{Measure and Hypothesis}

\begin{table}[t]
\centering
\caption{Post-questionnaire used in Study 2 to collect qualitative feedback on the usefulness of olfactory cues and navigation techniques.}
\label{tab:post-questionnaire}
\renewcommand{\arraystretch}{1.25}
\begin{tabular}{p{0.02\linewidth} p{0.9\linewidth}}
\toprule
\textbf{} & \textbf{Question} \\
\midrule
\textbf{Q1} &
\HDY{How was scent useful in performing the task? Were any specific scents (\textit{coffee}, \textit{fruity}, \textit{floral}, or \textit{pizza}) particularly memorable? Please explain why.} \\
\textbf{Q2} &
\HDY{Which navigation technique (\textit{Dash} or \textit{Teleportation}) did you prefer for completing the task? Please explain why.} \\
\textbf{Q3} &
\HDY{How would you describe your overall navigation experience with olfactory cues? What aspects did you find easy or difficult?} \\
\bottomrule
\end{tabular}
\vspace{-.4cm}
\end{table}

In Study 2, we employed the same evaluation metrics as in Study 1: Task Accuracy, SUS, NASA-TLX, and SSQ.
\HDY{At the end of the study, participants answered the following questions regarding their navigation experience with olfactory cues and the effectiveness of these cues in performing the task (Table~\ref{tab:post-questionnaire}).
}
    


To answer the Study 2 research question, we formulated the following hypotheses.

\begin{enumerate}[label=H\arabic*., noitemsep]\setcounter{enumi}{4}
    \item \textit{Vent2S} will yield higher accuracy than \textit{Vent1S}, as longer ventilation durations more effectively mitigate scent blending.

    \item Participants’ accuracy is expected to decrease as the movement sequence progresses from \textit{MV1} to \textit{MV4}, due to increasing scent blending across successive navigations within a trial.
    
    \item \textit{Vent2S} is expected to have a better scented navigation experience (i.e., higher SUS, lower NASA-TLX, and lower SSQ) than the \textit{Vent1S} condition, for the same reason as in H5.


    
  
\end{enumerate}

\subsection{Participants and Procedures}
We recruited 24 participants (22 male, 2 female, with a mean age of 26.2 years (range: 23–34, SD = 2.14). 
The same inclusion and exclusion criteria for participation were applied.
All participants had normal or corrected-to-normal vision (20/20) as well as the normal sense of smell.
The Meta Quest 2, olfactory device, and study environment were identical to those used in Study 1.
\HDY{None of the participants from Study 1 participated in Study 2.}


Study 2  took approximately 50 minutes. 
Its procedure was almost identical to that in Study 1.
Each participant performed 40 task trials and provided a total of 160 responses (2 navigation techniques $\times$  2 ventilation duration $\times$  4 answers in each trial $\times$ 10 repetitions). 
Participants received \$7 rewards for their participation.

\subsection {Results}

We used the same analytical procedures as in Study 1, calculating the average values, establishing 95\% CIs, and differences across conditions. Rather than presenting results for each condition, this section reports and compares results grouped by condition. 
\HDY{The detailed results are reported in Table~\ref{tab:study2_bootstrap_summary} of the supplementary material.}

\subsubsection{Task Accuracy}

\begin{figure}[t]
  \centering 
  \includegraphics[width=.95\linewidth]{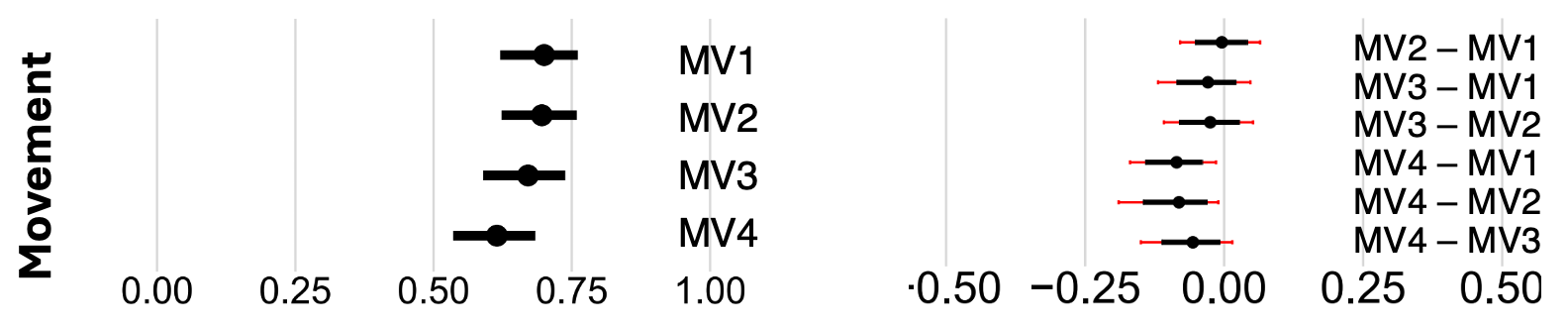}
  \vspace{-.2cm}
  \caption{Study 2 - Task accuracy by successive movement, with 95\% CI error bars.
   It shows the measurement average first and the pairwise comparison results next. 
  }
  \label{fig:study2_accuracy}
  \vspace{-.4cm}
\end{figure}

Regarding movement order, we found evidence of differences between MV1 (0.70, \KHC{\ciValues{CI=[0.62, 0.76]}}) and MV4 (0.61, \KHC{\ciValues{CI=[0.53, 0.68]}}), indicating that MV1 achieved higher accuracy than MV4 by 0.09, \KHC{\ciValues{CI=[0.01, 0.17]}}.  Similarly, MV2 (0.69, \KHC{\ciValues{CI=[0.62, 0.76]}}) showed higher accuracy than MV4 by 0.08, \KHC{\ciValues{CI=[0.01, 0.19]}}. The results are shown in Fig.~\ref{fig:study2_accuracy}
No other clear differences were observed between movement orders.  The accuracy for MV3 was 0.67, \KHC{\ciValues{CI=[0.59, 0.73]}}. The results partially support \textbf{H6}.

In contrast, we found no clear difference between Dash (0.69, \KHC{\ciValues{CI=[0.66, 0.73]}}) and Teleportation (0.67, \KHC{\ciValues{CI=[0.63, 0.70]}}).
Similarly, no clear difference was observed between Vent2S (0.69, \KHC{\ciValues{CI=[0.66, 0.73]}}) and Vent1S (0.67, \KHC{\ciValues{CI=[0.63, 0.70]}}), rejecting \textbf{H5}.


\subsubsection{\HDY{System Usability Scale} (SUS)}
We found no differences between Dash (79.52, \KHC{\ciValues{CI=[74.31, 83.93]}}) and Teleportation (80.63, \KHC{\ciValues{CI=[75.82, 84.73]}}), nor between Vent1S (80.91, \KHC{\ciValues{CI=[75.88, 85.22]}}) and Vent2S (79.25, \KHC{\ciValues{CI=[74.38, 83.52]}}). This rejects \textbf{H7}.

\subsubsection{NASA-TLX}
\textbf{Mental Workload:} Participants in Vent1S (7.89, \KHC{\ciValues{CI=[6.25, 9.60]}}) reported lower mental workload than those in Vent2S (8.85, \KHC{\ciValues{CI=[7.21, 10.60]}}), with a mean difference of 0.96, \KHC{\ciValues{CI=[0.08, 2.21]}}. However, we found no clear difference between Dash (8.46, \KHC{\ciValues{CI=[6.81, 10.17]}}) and Teleportation (8.29, \KHC{\ciValues{CI=[6.60, 10.04]}}).


\textbf{Physical Workload:}
Neither technique (Dash: 4.88, \KHC{\ciValues{CI=[3.42, 6.67]}}; Teleportation: 4.75, \KHC{\ciValues{CI=[3.25, 6.69]}}) nor ventilation time (Vent1S: 4.83, \KHC{\ciValues{CI=[3.31, 6.75]}}; Vent2S: 4.79, \KHC{\ciValues{CI=[3.38, 6.58]}}) showed a clear difference.


\textbf{Sense of Frustration:}
A lower value was reported for Vent2S (1.67, \KHC{\ciValues{CI=[0.94, 2.69]}}) compared to Vent1S (2.00, \KHC{\ciValues{CI=[1.17, 3.10]}}), with a difference of 0.33, \KHC{\ciValues{CI=[0.02, 0.81]}}. However, no clear difference between Dash (1.69, \KHC{\ciValues{CI=[0.95, 2.71]}}) and Teleportation (1.98, \KHC{\ciValues{CI=[1.17, 3.04]}}) was found.


% ...

\textbf{Performance:}
We found no differences between Dash (11.42, \KHC{\ciValues{CI=[9.75, 12.75]}}) and Teleportation (11.42, \KHC{\ciValues{CI=[9.90, 12.81]}}), nor between Vent1S (11.10, \KHC{\ciValues{CI=[9.58, 12.46]}}) and Vent2S (11.73, \KHC{\ciValues{CI=[10.08, 13.06]}}). 

\textbf{In summary,} we observed a compound effect.  Vent2S resulted in lower frustration but higher mental workload than Vent1S, while no differences were observed in physical workload or performance. Therefore, \textbf{H7} was only partially supported.


\begin{figure}[t]
  \centering 
  \includegraphics[width=.85\linewidth]{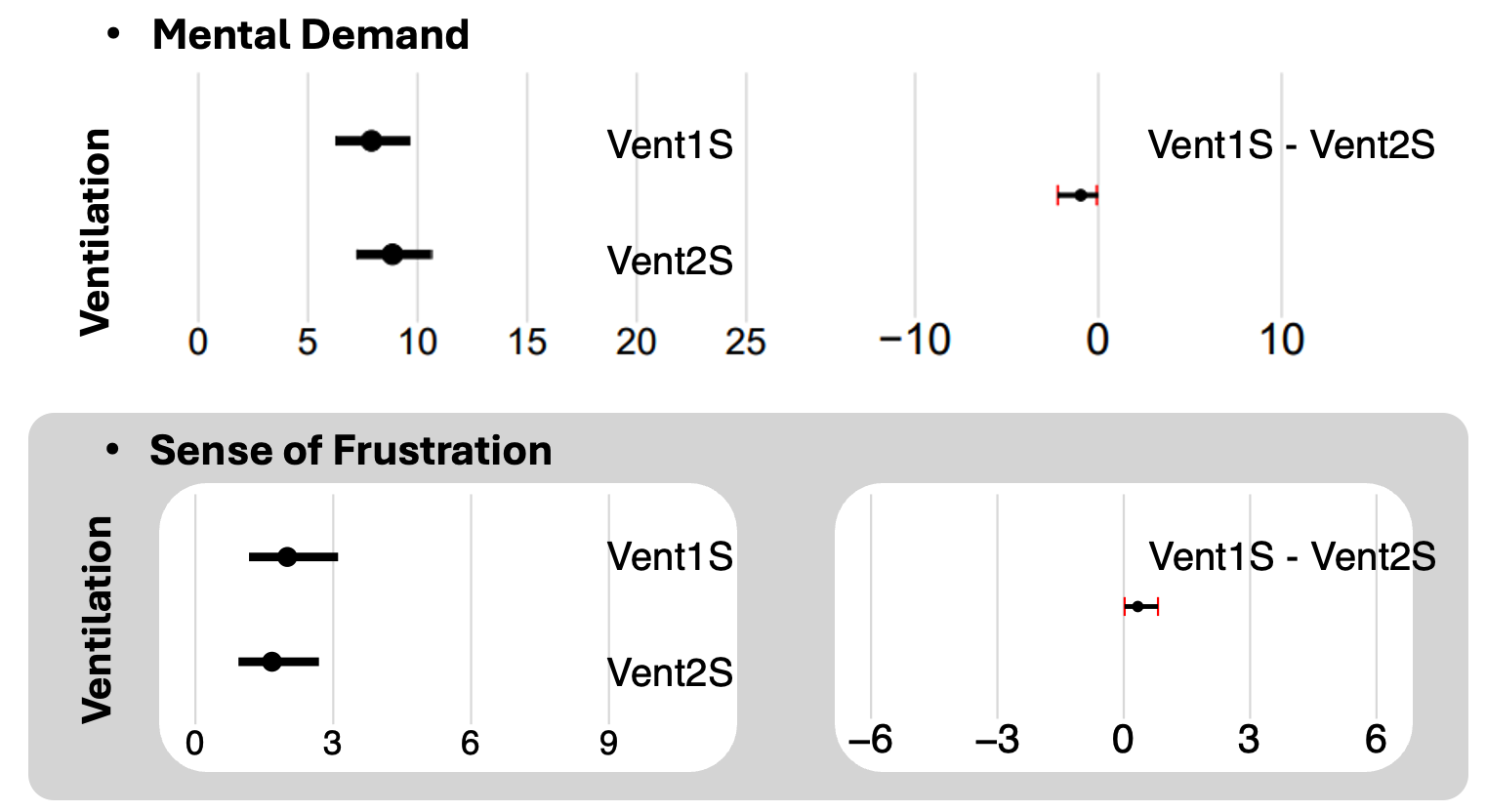}
  \caption{Study 2 - NASA-TLX Results (Mental Demans and Sense of Frustration) with 95\% CI error bars. 
  Each row shows the measurement average first and the pairwise comparison results next. 
 }
  \label{fig:study2_NASA}
\end{figure}

\subsubsection{\HDY{Simulator Sickness Questionnaire} (SSQ)}
\begin{figure}[t]
  \centering 
  \includegraphics[width=.9\linewidth]{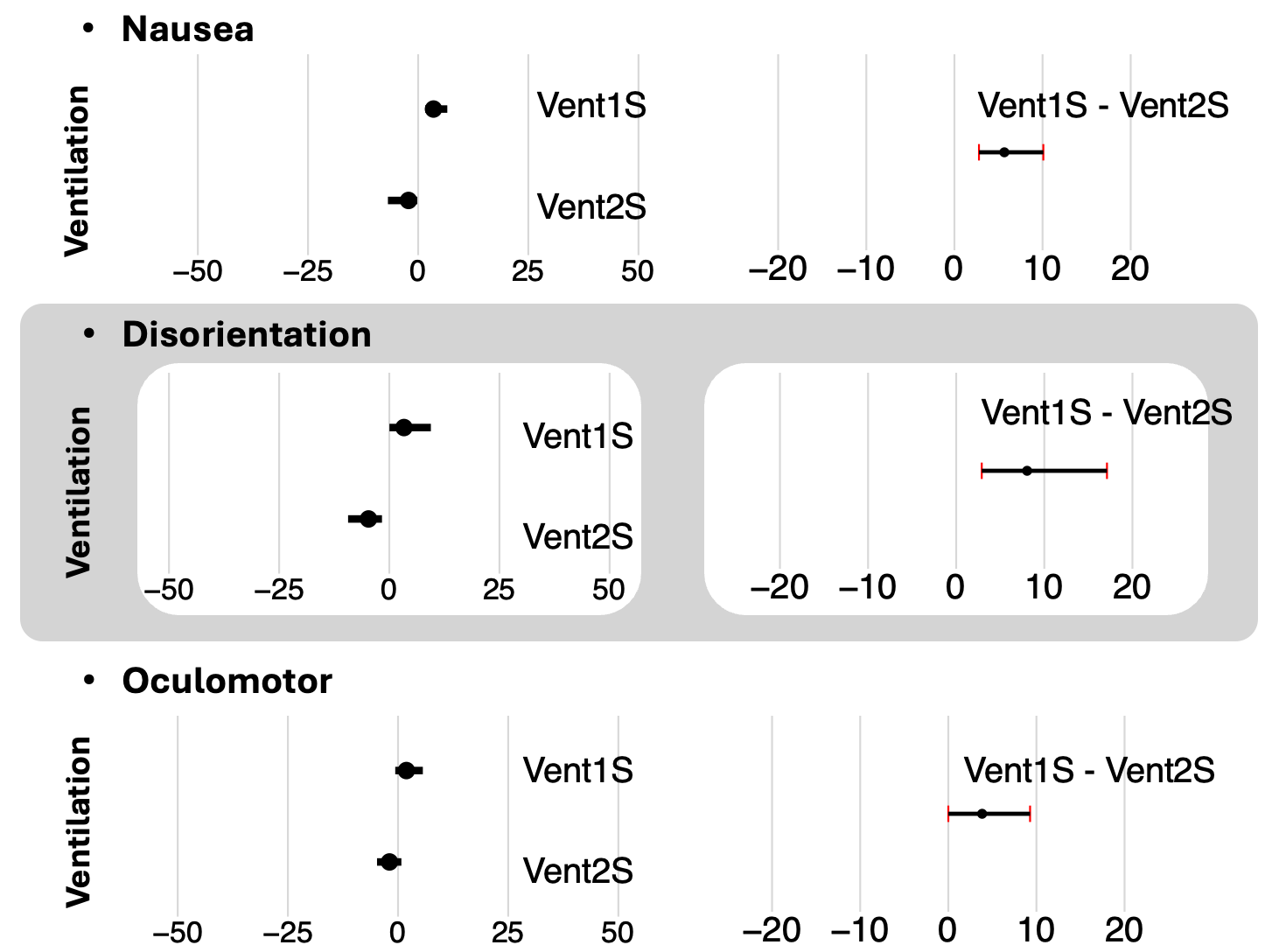}
  \caption{Study 2 - SSQ Survey Results with 95\% CI error bars. 
  Each row shows the measurement average first and pairwise comparison results next.
  The average results represent the difference between the post and baseline-measurement.  
  }
  \label{fig:study2_SSQ}
  \vspace{-.4cm}
\end{figure}

The results for each subscale are summarized in Fig.~\ref{fig:study2_SSQ}.

\textbf{Nausea:} 
Vent2S (-2.18, \KHC{\ciValues{CI=[-6.85, -0.20]}}) exhibited smaller changes than Vent1S (3.48, \KHC{\ciValues{[1.59, 6.61]}}), yielding a difference of 5.66, \KHC{\ciValues{CI=[2.78, 10.09]}}. However, no difference between Dash (-0.49, \KHC{\ciValues{CI=[-4.46, 2.43]}}) and Teleportation (1.79, \KHC{\ciValues{CI=[0.00, 3.88]}}) was found.

\textbf{Disorientation:}
Changes were smaller in Vent2S (-4.71, \KHC{\ciValues{CI=[-9.35, -1.66]}}) than in Vent1S (3.34, \KHC{\ciValues{CI=[0.03, 9.43]}}), with a difference of 8.05, \KHC{\ciValues{CI=[2.90, 17.11]}}.
However, no difference was observed between Dash (-1.74, \KHC{\ciValues{CI=[-5.94, 2.68]}}) and Teleportation (0.36, \KHC{\ciValues{CI=[-2.97, 5.36]}}).

\textbf{Oculomotor:} 
We also found a clear difference between Vent1S (1.90, \KHC{\ciValues{CI=[-0.63, 5.62]}}) and Vent2S (-1.94, \KHC{\ciValues{CI=[-4.76, 0.75]}}), with Vent2S showing a smaller change than Vent1S by 3.84, \KHC{\ciValues{CI=[0.00, 9.28]}}. However, Teleportation (0.79, \KHC{\ciValues{CI=[-1.69, 3.47]}}) and Dash (-0.83, \KHC{\ciValues{CI=[-2.64, 2.91]}}) yielded comparable changes.

\textbf{In summary,} Vent2S yielded smaller changes across all SSQ metrics, supporting H7.

\subsection{Study 2: Discussion}


\textbf{Adding visual motion cues to olfactory cues did not yield a clear improvement in recognition accuracy.}
In Study 2, Teleportation and Dash did not show a clear difference in task accuracy, consistent with the pattern observed in Study 1.
Participants' feedback revealed both benefits and drawbacks of the visual cues. 
Most participants (18 out of 24) reported that the visual cue helped them understand their current direction or position during navigation. 
Among those who found the visual cues helpful, six participants reported that it made the consecutive movements more noticeable, allowing them to better perceive the continuity of the navigation process. One participant noted that the visual cue made the movement feel slower than it actually was.  
More specifically, P11 commented that \textit{``Dash was much easier, and the sequence was clearer,'’} while P14 noted that \textit{`‘Dash felt like moving through the space, which made the time feel slower and made it easier to remember.''}

However, three participants mentioned that receiving both visual and olfactory cues simultaneously sometimes felt cognitively demanding. 
P4 commented that \textit{``The visual effect distracted me and made it harder to focus on the scent.''} and P20 said \textit{``Teleportation made it easier to perceive the store in order, and it felt cognitively easier due to the lower visual load.''}
These observations may help explain why adding visual motion cues to olfactory cues did not produce a clear accuracy advantage over Teleportation alone.

\textbf{As the number of navigations increased, later movements showed lower recognition accuracy. However, scent cues still appeared to support recognition during navigation.}
The Study 2 results showed that earlier movements (i.e., MV1 and MV2) achieved higher accuracy than MV4. 
One possible explanation is that olfactory cues may have been more perceptually distinct earlier in the sequence, whereas later movements may have been more affected by residual scent from previous navigations. However, as participants continued navigating, previous scents may have lingered in the environment and partially mixed with newly presented scents, making it more difficult to clearly distinguish between them.
As a result, scents encountered later in the navigation sequence may have become less perceptually distinct than those encountered earlier, which could have made it harder for participants to correctly identify the locations associated with those scents.
Nevertheless, considering the D results from Study 1, which showed an average task accuracy of about 30\%, the presence of scent cues may still have provided some support for participants in identifying the locations they had encountered during navigation. 

\HDY{
Participants' feedback indicated that stronger scents tended to be more memorable.
In particular, participants reported that the pizza scent was the most memorable; however, it was the least preferred scent. 
Additionally, participants reported that the coffee scent was generally easy to recognize, whereas the flower and orange scents were perceived as relatively weaker and were therefore mentioned as being less memorable.
}

Ventilation duration did not yield a clear difference in recognition accuracy. We did not observe a clear accuracy difference between Vent1S and Vent2S. One possible explanation is that the difference between the two conditions may have been limited in practice. In our implementation, the fan required approximately 0.5s to fully activate, reducing the effective ventilation time available in the shorter condition. As a result, the practical difference in scent-clearing between Vent1S and Vent2S may have been smaller than the nominal interval difference, which may help explain the absence of a clear performance difference. This may also suggest that, under our setup, even the shorter ventilation interval was practically adequate for maintaining recognition performance.

\textbf{A longer ventilation period was associated with smaller increases in simulator sickness-related measures.}
Study 2 showed that SSQ changes were generally smaller under the longer ventilation condition.
This pattern suggests that ventilation may help reduce discomfort during repeated scented navigation.
However, introducing a pause (i.e., 2.5s in Study 2) after every navigation event to allow for ventilation may reduce interaction efficiency and interrupt the flow of VR exploration.
These findings point to the potential value of ventilation, while also indicating the need for more effective scent-clearing strategies that better support continuous navigation.

We also observed that Vent2S resulted in a lower level of frustration compared to Vent1S. 
One possible explanation is that the longer ventilation duration may have made successive scents easier to distinguish during repeated navigation, which may have contributed to a more comfortable sensory experience and lower frustration.
In contrast, Vent2S showed a higher mental demand than Vent1S, contrary to our expectation.
One possible explanation is that the longer ventilation period introduced additional sensory changes that participants needed to process~\cite{wickens2008multiple}, such as airflow from the fans and the gradual dissipation of scents. 
Processing these sensory changes may have required additional perceptual effort, thereby increasing perceived mental demand. However, further controlled measurements are needed to clearly identify the mechanisms underlying this effect.

\section{Limitations and Future Work}
We acknowledge several study limitations.
One limitation of our study is the simple approach for scent exposure.
The intensity and duration of scent exposure could also play a crucial role in determining its effectiveness. 
As olfactory experiences are subjective and can vary significantly between individuals, allowing users to customize scent intensity may enhance their overall experience in VR ~\cite{Maggioni2020}. 
Ranasinghe et al.~\cite{ranasinghe2020exploring} and Javerliat et al.~\cite{javerliat2022nebula} attempted to control scent intensity by modulating the signal sent to the scent delivery fans. 
Although the actual intensity levels were not precisely quantified, their participants were able to successfully distinguish between three different intensity levels.  
Adopting a similar approach may enable limited control over scent intensity by adjusting diffusion levels.

\HDY{Another limitation concerns the ventilation mechanism of the current olfactory device. Although our results indicate that ventilation is a critical component for supporting olfactory experiences during navigation, its performance was not fully quantitatively validated in this work. Specifically, the prototype does not incorporate a sensor capable of measuring residual scent concentration or objectively verifying complete scent removal. Consequently, we were unable to evaluate the exact time required to eliminate scent or the effectiveness of the ventilation system under different operation condition. 
Future work should further evaluate the ventilation performance and investigate how different levels of ventilation efficiency affect user experiences in VR.
}

Next, the optimal duration for delivering olfactory cues during navigation requires further investigation.
This study set the navigation duration to 1.5 seconds, based on the reasoning outlined in Sec~\ref{sec_task}. 
However, in many VR applications, teleportation occurs within a few hundred milliseconds, shorter than the duration used in our study. 
Conversely, a navigation duration of 1.5 seconds may be too short for Dash, potentially limiting participants’ ability to clearly perceive and utilize both visual and olfactory cues during movement.
Moreover, it is also necessary to consider the presence and size of the vignetting effect~\cite{norouzi2018assessing, lim2024effects}.
To effectively integrate both visual and olfactory cues, future research should investigate the optimal navigation duration for Dash as well as the appropriate configuration of the vignetting effect to ensure sufficient visual information acquisition and an improved user experience. 
Furthermore, additional research is needed to identify effective scent ventilation strategies within short yet meaningful time intervals in order to support a more effective visuo-olfactory experience in VR.

In future work, we will investigate the impact of olfactory cues on a broader range of VR interaction techniques~\cite{al2018virtual}. 
\HDY{While this study employed relatively simple olfactory cues with two navigation techniques, richer olfactory interactions need to be explored. For example, continuous locomotion may require gradual modulation of scent intensity and dynamic blending of multiple scents to provide more realistic user experiences for transitions between virtual locations.
We will further investigate how olfactory cues can be integrated into other locomotion techniques, such as redirected walking~\cite{langbehn2017application}, and evaluate their effects on user experience and spatial awareness. This work will also involve developing more advanced olfactory devices capable of precise scent intensity control and effecient scent dispersion.
}

\section{Conclusion}

This work investigated how olfactory cues can be incorporated into teleportation-based VR navigation through a wearable olfactory prototype capable of emitting four distinct scents and providing active ventilation. Across two user studies, our results indicate that olfactory cues can support recognition of encountered locations during navigation, and that ventilation can benefit scent management and usability, particularly in single-navigation tasks. In repeated-navigation tasks, longer ventilation did not clearly improve recognition accuracy, but it was associated with smaller increases in simulator sickness-related measures and lower frustration. These findings suggest that olfactory cues can support recognition during VR navigation under the studied conditions, while also highlighting the role of ventilation and scent management in maintaining perceptual clarity and user comfort. 



\acknowledgments{
This research was supported by Culture, Sports and Tourism R\&D Program through the Korea Creative Content Agency grant funded by the Ministry of Culture, Sports and Tourism in 2026 (Project Number: RS-2026-25525669, Contribution Rate: 33\%), the Institute of Information \& Communications Technology Planning \& Evaluation (IITP) grant funded by the Korea government (MSIT) (RS-2026-25516382, 33\%), and the National Research Foundation of Korea (NRF) grant funded by the Korea government (MSIT) (RS-2026-25492756).}

\bibliographystyle{abbrv-doi}

\bibliography{0_references}

\newpage
\clearpage
\onecolumn

  \centering
  {\textsf{\huge Exploring the Effects of Olfactory Cues and Ventilation on Teleportation-based Navigation in VR }}
  \vskip 5pt
  \large \textsf{Supplemental Material}
  \vskip 10pt
\raggedright This document provides additional results that could not be included in the main paper due to space limitations.

\renewcommand{\figurename}{Fig.}
 \renewcommand\thefigure{\arabic{figure}}  
 \setcounter{figure}{0} 
\renewcommand{\thesection}{\Alph{section}}
\setcounter{section}{0}

\section{Study 1 Supplemental Material}

\subsection{Baseline Comparison Results}

\noindent
\begin{minipage}{\textwidth}
\captionsetup{type=table}
\centering
\caption{Bootstrapped means, 95\% confidence intervals (CIs), and pairwise differences for the Study 1 metrics are reported. Pairwise differences with 95\% CIs that do not include zero are considered statistically significant.}
\label{study1_all_results_baseline_comparison}
\centering
\setlength{\tabcolsep}{3pt}
\small
\resizebox{\textwidth}{!}{
\begin{tabular}{ll c c G c G c G c G}
\toprule
 & & \textbf{Baseline} & \multicolumn{4}{c}{\textbf{Teleportation}} & \multicolumn{4}{c}{\textbf{Dash}} \\
\cmidrule(r){4-7} \cmidrule(l){8-11}
\textbf{Measure} & \textbf{Subscale} 
& \textbf{Dash (D)} 
& \parbox[c][2em][c]{2cm}{\centering\textbf{Scent+Ventilation}\\\textbf{(TS+V)}}
& \parbox[c][2em][c]{2cm}{\centering\textbf{Diff}\\\textbf{(TS+V - D)}}
& \parbox[c][2em][c]{2cm}{\centering\textbf{Scent Only}\\\textbf{(TS-V)}}
& \parbox[c][2em][c]{2cm}{\centering\textbf{Diff}\\\textbf{(TS-V - D)}} 
& \parbox[c][2em][c]{2cm}{\centering\textbf{Scent+Ventilation}\\\textbf{(DS+V)}}
& \parbox[c][2em][c]{2cm}{\centering\textbf{Diff}\\\textbf{(DS+V - D)}} 
& \parbox[c][2em][c]{2cm}{\centering\textbf{Scent Only}\\\textbf{(DS-V)}} 
& \parbox[c][2em][c]{2cm}{\centering\textbf{Diff}\\\textbf{(DS-V - D)}} \\
\midrule
\textbf{Task Accuracy} 
 & - & 0.30 [0.24, 0.36] & 0.80 [0.73, 0.86] & \textbf{0.50 [0.40, 0.60]} & 0.67 [-0.58, 0.73] & \textbf{0.38 [0.23, 0.49]} & 0.77 [0.70, 0.84] & \textbf{0.47 [0.35, 0.58]} & 0.63 [0.55, 0.69] & \textbf{0.34 [0.22, 0.45]} \\
\midrule
\textbf{SUS} 
 & - & 61.13 [50.50, 72.00] & 86.50 [81.13, 90.75] & \textbf{25.38 [11.13, 42.25]} & 79.75 [73.75, 85.25] & \textbf{18.63 [7.65, 32.75]} & 85.75 [79.63, 90.75] & \textbf{24.63 [10.50, 41.34]} & 79.75 [70.00, 86.88] & \textbf{18.63 [6.63, 36.04]} \\
\midrule
\multirow{6}{*}{\textbf{NASA-TLX}} 
 & Mental Demand & 11.70 [8.95, 14.35] & 6.75 [4.90, 8.70] & \textbf{-4.95 [-7.80, -1.60]} & \nonsig{9.50 [7.10, 11.45]} & \nonsig{-2.20 [-6.90, 1.20]} & \nonsig{10.55 [8.25, 12.60]} & \nonsig{-1.15 [-4.45, 2.35]} & \nonsig{10.35 [8.05, 12.60]} & \nonsig{-1.35 [-5.30, 3.15]} \\
 & Physical Demand & 11.30 [8.55, 13.90] & 5.95 [4.15, 8.00] & \textbf{-5.35 [-9.05, -2.00]} & \nonsig{7.75 [5.45, 10.40]} & \nonsig{-3.55 [-7.15, 0.76]} & \nonsig{6.55 [4.10, 9.35]} & \nonsig{-4.75 [-9.30, 0.25]} & 6.15 [3.89, 9] & \textbf{-5.15 [-9.90, -0.20]} \\
 & Frustration & 2.70 [1.25, 5.01] & \nonsig{0.70 [0.20, 1.55]} & \nonsig{-2.00 [-5.30, 0.00]} & \nonsig{2.70 [1.20, 5.10]} & \nonsig{0.00 [-2.80, 2.95]} & \nonsig{1.35 [0.60, 2.85]} & \nonsig{-1.35 [-4.42, 0.35]} & \nonsig{2.4 [1.3, 4]} & \nonsig{-0.3 [-3.05, 2.20]} \\
 & Performance & 14.40 [11.45, 16.55] & 4.80 [3.10, 6.55] & \textbf{-9.60 [-13.85, -4.70]} & 9.45 [7.25, 11.70] & \textbf{-4.95 [-8.75, -1.40]} & 7.70 [5.60, 9.65] & \textbf{-6.70 [-10.80, -2.20]} & 7.25 [5.45, 8.65] & \textbf{-7.15 [-10.27, -3.75]} \\
\midrule
\multirow{3}{*}{\textbf{SSQ}} 
 & Nausea & 0.95 [0.25, 1.85] & \nonsig{0.50 [-0.35, 1.55]} & \nonsig{-0.45 [-2.30, 0.95]} & \nonsig{0.90 [0.15, 1.65]} & \nonsig{-0.05 [-1.25, 1.40]} & \nonsig{-0.30 [-2.15, 0.50]} & \nonsig{1.25 [-3.65, 0.20]} & \nonsig{1.05 [0.45, 1.70]} & \nonsig{0.1 [-0.90, 1.45]} \\
 & Disorientation & 0.85 [0.15, 1.60] & \nonsig{0.40 [-0.40, 1.90]} & \nonsig{-0.45 [-2.00, 1.37]} & \nonsig{0.20 [-0.25, 0.65]} & \nonsig{-0.65 [-1.60, 0.25]} & \nonsig{-0.10 [-1.25, 0.83]} & \nonsig{-0.95 [-2.75, 0.40]} & \nonsig{1.05 [0.5, 1.75]} & \nonsig{0.2 [-0.75, 1.70]} \\
 & Oculomotor & 1.55 [0.25, 3.15] & \nonsig{1.75 [0.40, 3.45]} & \nonsig{0.20 [-2.48, 2.65]} & \nonsig{1.45 [0.50, 2.65]} & \nonsig{-0.10 [-1.68, 1.80]} & \nonsig{0.65 [-0.65, 1.90]} & \nonsig{-0.90 [-2.35, 1.00]} & \nonsig{1.80 [0.45, 3.40]} & \nonsig{0.25 [-1.15, 2.10]} \\
\bottomrule
\addlinespace
\multicolumn{8}{l}{\textit{Note.} For Differences, intervals excluding 0 indicate statistically clear effects and are highlighted in \textbf{bold}.} \\
\multicolumn{8}{l}{SSQ values represent the mean change ($\Delta$) from the baseline pre-measurement.}
\end{tabular}}
\end{minipage}

\subsection{Group Analysis Results}

\noindent
\begin{minipage}{\textwidth}
\captionsetup{type=table}
\centering
\caption{Bootstrapped means, 95\% confidence intervals (CIs), and pairwise differences for the Study 1 metrics are reported. Pairwise comparisons evaluate the effects of navigation type (TS vs. DS) and ventilation (S+V vs. S-V). Pairwise differences whose 95\% CIs exclude zero are considered statistically significant.}
\label{study1_all_results_group_analysis}
\centering
\setlength{\tabcolsep}{3pt}
\resizebox{\textwidth}{!}{
\begin{tabular}{llcccccc}
\toprule
 &  & \multicolumn{3}{c}{\textbf{Navigation Technique (Scent)}} & \multicolumn{3}{c}{\textbf{Ventilation Presence}} \\
\cmidrule(r){3-5} \cmidrule(l){6-8}
\textbf{Measure} & \textbf{Subscale} &  \textbf{Teleport (TS)} & \textbf{Dash (DS)} & \cellcolor{gray!15}\textbf{Difference (TS$-$DS)} & \textbf{With Vent (S+V)} & \textbf{No Vent (S-V)} & \cellcolor{gray!15}\textbf{Difference (S+V$-$S-V)} \\
\midrule
\textbf{Task Accuracy} 
 & - & \nonsig{0.74 [0.68, 0.78]} & \nonsig{0.70 [0.64, 0.75]} & \cellcolor{gray!15}\nonsig{0.03 [-0.04, 0.11]} & 0.79 [0.73, 0.83] & 0.65 [0.59, 0.70] & \cellcolor{gray!15}\textbf{0.13 [0.06, 0.20]} \\
\midrule
\textbf{SUS} 
 & - & \nonsig{83.13 [79.10, 86.69]} & \nonsig{82.75 [77.00, 87.31]} & \cellcolor{gray!15}\nonsig{0.38 [-4.00, 4.94]} & 
 86.13 [82.19, 89.50] & 
 79.75 [74.00, 84.25] & \cellcolor{gray!15}\textbf{6.38 [1.88, 12.38]} \\
\midrule
\multirow{6}{*}{\textbf{NASA-TLX}} 
 & Mental Demand & 8.13 [6.63, 9.58] & 10.45 [8.83, 11.98] & \cellcolor{gray!15}\textbf{-2.33 [-4.80, -0.35]} & \nonsig{8.65 [7.13, 10.20]} & \nonsig{9.93 [8.28, 11.48]} & \cellcolor{gray!15}\nonsig{-1.28 [-3.45, 1.13]} \\
 & Physical Demand  & \nonsig{6.85 [5.38, 8.50]} & \nonsig{6.35 [4.68, 8.28]} & \cellcolor{gray!15}\nonsig{0.50 [-2.08, 2.98]} & \nonsig{6.25 [4.73, 7.90]} & \nonsig{6.95 [5.30, 8.83]} & \cellcolor{gray!15} \nonsig{-0.70 [-3.20, 1.73]} \\
& Frustration & \nonsig{1.70 [0.88, 3.00]} & \nonsig{1.88 [1.18, 2.95]} & \cellcolor{gray!15} \nonsig{-0.18 [-1.50, 1.03]} & 1.03 [0.53, 1.83] & 2.55 [1.55, 3.85] & \cellcolor{gray!15} \textbf{-1.53 [-3.28, -0.13]} \\
 & Performance  & \nonsig{7.13 [5.60, 8.75]} & \nonsig{7.48 [6.13, 8.73]} & \cellcolor{gray!15}\nonsig{-0.35 [-2.50, 1.99]} & \nonsig{6.25 [4.83, 7.68]} & \nonsig{8.35 [6.93, 9.78]} & \cellcolor{gray!15}\nonsig{-2.10 [-4.60, 0.18]} \\

\midrule
\multirow{3}{*}{\textbf{SSQ}} 
 & Nausea & \nonsig{0.70 [0.10, 1.33]} & \nonsig{0.38 [-0.60, 0.90]} & \cellcolor{gray!15}\nonsig{-0.25 [-1.28, 0.58]} & 0.10 [-0.85, 0.75] & 0.98 [0.48, 1.48] & \cellcolor{gray!15}\textbf{-0.88 [-2.03, -0.13]} \\
 & Disorientation  & \nonsig{0.30 [-0.18, 1.08]} 
 & \nonsig{0.48 [-0.23, 1.05]} & \cellcolor{gray!15}\nonsig{-0.55 [-1.40, 0.28]} & \nonsig{0.15 [-0.55, 0.98]} & \nonsig{0.63 [0.25, 1.08]} & \cellcolor{gray!15}\nonsig{-0.48 [-1.33, 0.60]} \\
 & Oculomotor  & \nonsig{1.60 [0.75, 2.58]} & \nonsig{1.23 [0.28, 2.25]} & \cellcolor{gray!15}\nonsig{0.05 [-1.35, 1.35]} & \nonsig{1.20 [0.25, 2.29]} & \nonsig{1.63 [0.78, 2.60]} & \cellcolor{gray!15}\nonsig{-0.43 [-1.53, 0.78]} \\
\bottomrule
\addlinespace
\multicolumn{8}{l}{\textit{Note.} For Differences, intervals excluding 0 indicate statistically clear effects and are highlighted in \textbf{bold}.} \\
\multicolumn{8}{l}{SSQ values represent the mean change ($\Delta$) from the baseline pre-measurement.}
\end{tabular}}
\end{minipage}

\section{Study 2 Supplemental Material}

\subsection{Study 2 results table}


\noindent
\begin{minipage}{\textwidth}
\captionsetup{type=table}
\centering
\caption{Summary of Bootstrapped Means and 95\% Confidence Intervals (CIs) for Study 2 Metrics are reported. Pairwise differences whose 95\% CIs exclude zero are considered statistically significant.}
\label{tab:study2_bootstrap_summary}
\centering
\small
\begin{tabular}{llcccccc}
\toprule
 & & \multicolumn{3}{c}{\textbf{Navigation Technique}} & \multicolumn{3}{c}{\textbf{Ventilation Duration}} \\
\cmidrule(r){3-5} \cmidrule(l){6-8}
\textbf{Measure} & \textbf{Subscale} & \textbf{Dash} & \textbf{Teleportation} & \cellcolor{gray!15}\textbf{Difference}  & \textbf{Vent1S} & \textbf{Vent2S} & \cellcolor{gray!15}\textbf{Difference}\\
\midrule
\textbf{Task Accuracy} 
 & - & \nonsig{0.69 [0.66, 0.73]} & \nonsig{0.67 [0.63, 0.70]} & \cellcolor{gray!15}\nonsig{-0.03 [-0.06, 0.01]} & \nonsig{0.67 [0.63, 0.70]} & \nonsig{0.69 [0.66, 0.73]} & \cellcolor{gray!15}\nonsig{-0.02 [-0.06, 0.01]}\\
\midrule
\textbf{SUS} 
 & - & \nonsig{79.52 [74.31, 83.93]} & \nonsig{80.63 [75.82, 84.73]} & \cellcolor{gray!15}\nonsig{1.11 [-0.97, 3.27]} & \nonsig{80.91 [75.88, 85.22]} & \nonsig{79.25 [74.38, 83.52]} & \cellcolor{gray!15}\nonsig{1.67 [-0.42, 3.99]}\\
\midrule
\multirow{6}{*}{\textbf{NASA-TLX}} 
 & Mental Demand & \nonsig{8.46 [6.81, 10.17]} & \nonsig{8.29 [6.60, 10.04]} & \cellcolor{gray!15}\nonsig{-0.17 [-1.15, 1]} & 7.89 [6.25, 9.60] & 8.85 [7.21, 10.60]  & \cellcolor{gray!15}\textbf{-0.96 [-2.20, -0.08]} \\
 & Physical Demand & \nonsig{4.88 [3.42, 6.67]} & \nonsig{4.75 [3.25, 6.69]}  & \cellcolor{gray!15}\nonsig{-0.13 [-0.79, 0.70]} & \nonsig{4.83 [3.31, 6.75]} & \nonsig{4.79 [3.38, 6.58]}  & \cellcolor{gray!15}\nonsig{0.04 [-0.41, 0.54]} \\
& Frustration & \nonsig{1.69 [0.95, 2.71]} & \nonsig{1.98 [1.17, 3.04]}  & \cellcolor{gray!15}\nonsig{0.29 [-0.23, 0.98]} & 2.00 [1.17, 3.10] & 1.67 [0.94, 2.69]  & \cellcolor{gray!15}\textbf{0.33 [0.02, 0.81]} \\
 & Performance & \nonsig{11.42 [9.75, 12.75]} & \nonsig{11.42 [9.90, 12.81]}  & \cellcolor{gray!15}\nonsig{0 [-1.35, 1.46]} & \nonsig{11.10 [9.58, 12.46]} & \nonsig{11.73 [10.08, 13.06]}  & \cellcolor{gray!15}\nonsig{-0.62 [-1.81, 0.67]} \\
\midrule
\multirow{3}{*}{\textbf{SSQ}} 
 & Nausea & \nonsig{-0.49 [-4.46, 2.43]} & \nonsig{1.79 [0.00, 3.88]}  & \cellcolor{gray!15}\nonsig{2.29 [-1.39, 6.61]} & 3.48 [1.59, 6.61] & -2.18 [-6.85, -0.20]  & \cellcolor{gray!15}\textbf{5.66 [2.78, 10.09]} \\
 & Disorientation & \nonsig{-1.74 [-5.94, 2.68]} & \nonsig{0.36 [-2.97, 5.36]}  & \cellcolor{gray!15}\nonsig{2.10 [-3.33, 9.28]} & 3.34 [0.03, 9.43] & -4.71 [-9.35, -1.66]  & \cellcolor{gray!15}\textbf{8.05 [2.90, 17.11]} \\
 & Oculomotor & \nonsig{-0.83 [-2.64, 2.91]} & \nonsig{0.79 [-1.69, 3.47]}  & \cellcolor{gray!15}\nonsig{1.63 [-2.99, 6.31]} & 1.90 [-0.63, 5.62] & -1.94 [-4.76, 0.75]  & \cellcolor{gray!15}\textbf{3.83 [0.00, 9.28]} \\
\bottomrule
\addlinespace
\multicolumn{6}{l}{\textit{Note.} For Differences, intervals excluding 0 indicate statistically clear effects and are highlighted in \textbf{bold}.} \\
\multicolumn{6}{l}{SSQ values represent the mean change ($\Delta$) from the baseline pre-measurement.}
\end{tabular}
\end{minipage}
\end{document}